\documentclass[final,authoryear,12pt,fleqn]{elsarticle}
\usepackage{natbib}
\usepackage{amssymb, amsmath}
\usepackage{amsfonts}
\usepackage{amsthm}
\usepackage{ushort}
\usepackage{graphics}
\usepackage{bm}
\usepackage{float}
\usepackage{appendix}
\usepackage{comment}
\usepackage[multiple]{footmisc}
\usepackage{rotating}
\usepackage{subcaption}
\usepackage{booktabs}
\usepackage{threeparttable}
\usepackage{pdflscape}
\usepackage{mathpazo}
\usepackage{setspace}
\usepackage[margin=1in]{geometry}

\newtheorem{theorem}{Theorem}

\newtheorem{corollary}{Corollary}
\theoremstyle{definition}
\newtheorem{assumption}{Assumption}[section]

\begin{document}
\bibpunct{(}{)}{,}{a}{}{,}
\bibliographystyle{ecta}
\numberwithin{equation}{section}
\numberwithin{lemma}{section}
\numberwithin{corollary}{section}
\numberwithin{remark}{section}
\numberwithin{theorem}{section}
\numberwithin{proposition}{section}
\def\sym#1{\ifmmode^{#1}\else\(^{#1}\)\fi}
\makeatletter
\def\ps@pprintTitle{%
 \let\@oddhead\@empty
 \let\@evenhead\@empty
 \def\@oddfoot{}%
 \let\@evenfoot\@oddfoot}
\makeatother
\begin{frontmatter}
\title{{\sc \LARGE{Multiple Structural Breaks in Interactive Effects Panel Data and the Impact of Quantitative Easing on Bank Lending}}}

\author[a]{Jan Ditzen}
\author[b]{Yiannis Karavias\corref{cor1}}
\author[c]{Joakim Westerlund}
\cortext[cor1]{Corresponding author. E-mail: i.karavias@bham.ac.uk. Department of Economics, University of Birmingham, Edgbaston, B15 2TT, Birmingham, UK.
The authors would like to thank Theologos Dergiades, Hashem Pesaran, Francesco Ravazzolo, Barbara Rossi,  and seminar participants at the Free University of Bozen-Bolzano and the Universities of Birmingham and Sheffield for many valuable comments and suggestions. The authors would like to thank Theologos Dergiades particularly for creating Figures 1 and 2. 
Ditzen acknowledges financial support from Italian Ministry MIUR under the PRIN project Hi-Di NET - Econometric Analysis of High Dimensional Models with Network Structures in Macroeconomics and Finance (grant 2017TA7TYC). Westerlund would like to thank the Knut and Alice Wallenberg Foundation for financial support through a Wallenberg Academy Fellowship.}
\address[a]{Free University of Bozen-Bolzano, Bozen, ,Italy}
\address[b]{Department of Economics, University of Birmingham, Birmingham, B152TT, UK}
\address[c]{Department of Economics, Lund University, Lund, SE--221 00, Sweden \\ and Deakin University, Melbourne, Victoria 3125, Australia}

\begin{abstract}
This paper develops a new toolbox for multiple structural break detection in panel data models with interactive effects. The toolbox includes tests for the presence of structural breaks, a break date estimator, and a break date confidence interval. The new toolbox is applied to a large panel of US banks for a period characterized by massive quantitative easing programs aimed at lessening the impact of the global financial crisis and the COVID--19 pandemic. The question we ask is: Have these programs been successful in spurring bank lending in the US economy? The short answer turns out to be: ``No''.
\end{abstract}

\end{frontmatter}

\section{Introduction}\label{section::intro}

Accounting for structural change has always been an important issue in economics and elsewhere. However, because of major events such as the 2007--2008 global financial crisis, the 2007--2010 subprime mortgage crisis, the 2016 Brexit referendum, the 2020 COVID--19 outbreak, and the 2022 war in Ukraine, interest has recently intensified.

The time series literature concerned with the estimation and testing for breaks is huge, and there is by now considerable accumulated empirical evidence of breaks in all kinds of economic relationships, especially in macroeconomics and finance (see, for example, Stock and Watson, 1996, 2003). Most of this evidence is based on econometric techniques requiring that there is (at most) one break (see, for example, Andrews, 1993). However, there are also techniques that allow for an unknown number of breaks, which is typically the most relevant scenario in practice. The single most important contribution in this area is Bai and Perron (1998), ``BP98'' henceforth, who develop a complete toolbox for testing and dating multiple (discrete) breaks in linear time series regression models. The toolbox includes (i) a number of tests for the presence of breaks, including a sequential test procedure to estimate the number of breaks, (ii) a breakpoint estimator, and (iii) a breakpoint confidence interval. The toolbox is widely applicable, it is computationally attractive, and it is readily available in many software programmes, such as GAUSS, EViews, MATLAB, R and most recently Stata (see Ditzen et al., 2022). It is therefore very popular. This is true not only among empirical economists but also among econometricians for which the BP98 study has provided a launching pad for countless extensions (see the 2021 Journal of Econometrics special issue in honor of Pierre Perron for overviews of some of these works). Even today, more then two decades after its publication, the BP98 toolbox remains the workhorse of the literature. It has therefore stood the test of time.

Panel data relationships are particularly susceptible to breaks, because of the number of series they contain. In fact, if one admits to the possibility that single time series may be subject to breaks, the likelihood of having at least one breaking series in a panel will by construction increase as more series are added. This fact is by now well understood in the econometric literature, and it is not difficult to find empirical evidence in its support (see Antoch et al., 2019, Boldea et al., 2020, Kaddoura and Westerlund, 2022, Karavias et al., 2022, and Zhu et al., 2020, to mention a few). Yet, it is not until recently that tools designed to estimate and test for breaks in panel data regression models have become available to practitioners. Most of these tools are constrained to single break environments (see, for example, Antoch et al., 2019, Baltagi et al., 2016, 2017, Karavias et al., 2022, and Zhu et al., 2020). Boldea et al. (2020), Kaddoura and Westerlund (2022), and Qian and Su (2016) allow for multiple breaks, and are therefore more general in this regard. However, this generality comes at a cost in terms of the allowable unobserved heterogeneity, which is important as unattended heterogeneity can be mistaken for breaks (and vice versa). In particular, while Qian and Su (2016) allow for individual fixed effects, Boldea et al. (2020), and Kaddoura and Westerlund (2022) assume that the unobserved heterogeneity is made up of a special type of interactive random effects that basically can be ignored in the estimation. Li et al. (2016) allow for more general interactive effects and multiple breaks that are dealt with using a version of the (group fused) least absolute shrinkage and selection operator (LASSO) method. As far as we are aware, this is the most general approach available at the moment. The use of the LASSO does, however, make for a rather complicated estimation problem that is not only nonlinear but that also involves a number of tuning parameters.\footnote{In the LASSO of Li et al. (2016) researchers need to specify a penalty tuning parameter, an adaptive weight, a shrinking threshold, and the number of factors. As is well known, performance can be very sensitive to choices of this type. Some can be made in a data-driven fashion using information criteria but then the use of such criteria in turn requires other choices, including a penalty scaling factor, the penalty itself and a suitable grid for the parameter search. Moreover, since the LASSO requires numerical optimization, the use of data-driven tuning parameter selection makes the procedure computationally very burdensome. The same is true for the LASSO considered by Kaddoura and Westerlund (2022).} It has therefore not received much attention in the applied literature.

In the present paper we take the above observations as a source of inspiration for developing the first fully fledged panel extension of BP98's time series toolbox, which as already pointed out provides a natural starting point.\footnote{The paper that is closest to ours in terms of the techniques used is that of Karavias et al. (2022). As already pointed out, however, they only allow for one break, which is unlikely to be enough in most scenarios of empirical relevance. Moreover, most of their theoretical results assume that $T$ is fixed, which is very different from the large-$T$ theory provided in the present paper. Their empirical application to the relationship between stock returns and COVID-19 is also completely different from ours. The two papers are therefore clearly distinct.} The model that we consider for this purpose, laid out in Section \ref{section::modelass}, is very general and allows not only multiple structural breaks but also interactive effects.\footnote{In this paper, we follow BP98 and assume that breaks are discrete. This is not necessary, though, and there are papers that consider smooth breaks (see, for example, Chen and Huang, 2018, and Feng et al., 2017). Most economic data sets do, however, have relatively low frequency of observation, which means that even if breaks are smooth they will appear as discrete. Discrete breaks are therefore empirically highly relevant, which is presumably also one of the reasons for why they are so popular in applied work.} As we explain in detail in Section \ref{section::toolbox}, in contrast to Li et al. (2016), the estimation of the interactive effects, the breaks and the model parameters is not carried out jointly but sequentially in two steps, which enables us to separate the breaks from the effects. We begin by estimating and removing the interactive effects, which is done using a version of the so-called ``common correlated effects'' (CCE) approach, which is simple to implement, computationally fast and with good small-sample properties. With the interactive effects gone, in the second step we apply a panel version of BP98's toolbox to the resulting cleaned up regression. Our theoretical results, also reported in Section \ref{section::toolbox}, show that the toolbox is asymptotically valid, a result that is verified in small samples in a Monte Carlo study reported in the online appendix.

While the US economy has been struck by several major events in the past 15 years, the 2007--2008 global financial crisis and the 2020 COVID--19 outbreak have been particularly disruptive. In both cases the Federal Reserve's policy response was the use of a series of large-scale asset purchase programs, or ``quantitative easing'' (QE) rounds, whose usefulness to this date is debated. In Section \ref{section::appl}, we evaluate whether these rounds achieved their primary goal of spurring credit flow in the economy. The idea here is that effective QE policies would cause breaks in the banks' lending behaviour, which can be detected using the proposed break toolbox. The attractiveness of this methodology over the otherwise so common difference-in-difference (DiD) approach is threefold. First, both the timing of the breaks and their number can be determined from the data. This is important because the QE rounds differed in policy mix, duration and magnitude, and there were also other policy actions such as maturity extension programs, mid-cycle expansions and taperings. It is therefore not clear where the effects of these rounds begin and end, or if they are absent altogether. Second, the toolbox is robust to the presence of interactive effects. This is important because many policy actions were taken based on current and anticipated macroeconomic conditions, which if not properly accounted for may well be mistaken for policy-induced breaks. Third, by identifying policy through breaks, we avoid the usual problem of having to define suitable treatment and control groups, which in the current context is non-trivial, as all banks are effected by QE to some extent.

The sample that we use covers 3,557 banks across 64 quarters, from 2005Q3 to 2021Q3. According to the results, bank lending has suffered from a number of breaks that can be attributed to the Federal Reserve's QE interventions. However, a majority of these did not have the intended expansionary effect, but rather banks took the opportunity to build up reserves in order to meet their capitalisation requirements. The main exception occurs towards the end of the sample, a period that includes the COVID--19 outbreak. At this point in time, the health of the banking sector had improved substantially, and the subsequent interventions were large enough for banks to be able to expand lending while at the same time cover their capitalisation needs. The net effect over the whole sample period is that lending has increased, but just barely.

Section \ref{section::concl} concludes. All proofs and theoretical results of secondary nature are provided in the online appendix, which also contains the results of a small-scale Monte Carlo study. In a companion paper, Ditzen et al. (2022) present a new Stata command, called \texttt{xtbreak}, for implementing the new toolbox. The command uses the dynamic programming algorithm presented in Section \ref{section::breakestim} to efficiently estimate the break dates. The algorithm requires at most $ O(T^2) $ least-squares operations for any number of breaks.

\section{Model and assumptions}\label{section::modelass}

\subsection{Model}\label{section::model}

Consider the following panel data model with $N$ cross-sectional units, $T$ time periods and $k$ structural breaks:
\begin{align}
y_{i,t}&=x_{i,t}'\beta+w_{i,t}'\delta_1+e_{i,t} \;\; \text{for} \;\; t=1,...,T_1,\notag\\
y_{i,t}&=x_{i,t}'\beta+w_{i,t}'\delta_2+e_{i,t} \;\; \text{for} \;\; t=T_1+1,...,T_2,\notag\\
&\vdots\notag\\
y_{i,t}&=x_{i,t}'\beta+w_{i,t}'\delta_{k+1}+e_{i,t}  \;\; \text{for} \;\; t=T_{k}+1,...,T,\label{eq:y}
\end{align}
where $T_{1},...,T_k$ are the dates of the $k$ structural breaks and where we additionally define $T_0=0$ and $T_{k+1}=T$. The dependent variable $y_{i,t}$ and the error $e_{i,t}$ are scalars, the regressors $x_{i,t}$ and $w_{i,t}$ are $p_x \times 1$ vectors and $p_w \times 1$ vectors, respectively, and $\beta$ and $\delta_1,...,\delta_{k+1}$ are conformable vectors of coefficients. The coefficients of $x_{i,t}$ are unaffected by the breaks, while those of $w_{i,t}$ are affected by the breaks. It is possible that all coefficients break, in which case we define $x_{i,t}'\beta = 0$. It is also possible that different coefficients break at different times by allowing subsets of $\delta_j$ to remain constant across regimes. We follow the bulk of the previous literature, and assume that $\beta$ and $\delta_1,...,\delta_{k+1}$ are equal across the cross-section. One way to relax this assumption is to follow Pesaran (2006), and to assume that the coefficients are random but with common means. This will not affect the validity of our toolbox, which can then be considered as estimating the mean of the individual coefficients. Random coefficients will, however, come at a cost of other restrictive conditions, more complicated proofs and relatively slow rates of convergence (see Westerlund, 2019). It will therefore not be considered here. The break dates are also assumed to be common across the cross-section, which is again standard. This condition is reasonable in the type of low-frequency applications considered in Section \ref{section::appl}. Again, if each series has its own breakpoints, the toolbox developed here can be considered as estimating the mean of the individual breakpoints (see Bai, 2010). 

In the empirical analysis of Section \ref{section::appl}, $y_{i,t} $ is one of three measures of lending by bank $i$ in quarter $t$. The regressors in $w_{i,t}$, whose coefficients are allowed to be breaking, are bank holdings in securities believed to be affected by QE. The regressors included in $x_{i,t}$ are control variables like returns on assets, total assets, equity, and cost of deposits, among others.

Unobserved heterogeneity in $e_{i,t}$ is allowed through the following interactive effects specification:
\begin{equation}
	e_{i,t}=f_t'\gamma_i+\varepsilon_{i,t},
	\label{eq:e}
\end{equation}
where $f_t$ is a $m\times 1$ vector of unobserved common factors with $\gamma_{i}$ being the associated vector of factor loadings, and $\varepsilon_{i,t}$ is an idiosyncratic error. The interactive effects are here given by $f_t'\gamma_{i}$. This specification is very general and nests many extant specifications. For example, if $m=1$ and $f_t=1$, then $ f_t'\gamma_{i}=\gamma_{i}$, which is the usual one-way error components model, whereas if $m=2$, $f_t=(1,\nu_t)'$ and $\gamma_{i}=(\xi_{i},1)'$, then $f_t'\gamma_{i}=\xi_{i}+\nu_t$, which is the usual two-way error components model. In our paper, $m$ and the form of $f_t'\gamma_{i}$ is unspecified. Because they are common to all cross-sectional units, the factors are a source of strong cross-section dependence.

The model in \eqref{eq:w} does not allow the breaks to affect the loadings in $\gamma_{i}$. This condition is not necessary. Breaks in $ \gamma_i$ that take place at the same dates as those in $ \delta_j $ can be permitted without affecting any of our results. Breaks that only affect $ \gamma_i $ or that take place at different dates than those in $ \delta_j $ can also be accommodated but only at a cost. In particular, since a model with one factor and a breaking loading can be written equivalently as a model without break but with two factors (see, for example, Breitung and Eickmeier, 2011), unattended breaks in the loadings will cause the number of factors to increase, and as we explain in detail in Section \ref{section::assump} the number of factors that we can accommodate is limited by the number of regressors. Unattended breaks can therefore lead to too many factors. In the empirical analysis of Section \ref{section::appl}, we explain how to accommodate observed common factors with breaking loadings.

In many empirical applications, it is likely that $f_t$ is correlated with the regressors. In order to capture this, we assume that
\begin{align}
x_{i,t}&=\Gamma_{x,i}' f_t+u_{x,i,t},\label{eq:x}\\
w_{i,t}&=\Gamma_{w,i}' f_t+u_{w,i,t},\label{eq:w}
\end{align}
where the factor loading matrices $\Gamma_{x,i}$ and $\Gamma_{w,i}$ are $m\times p_x$ and $m\times p_w$, respectively, while the idiosyncratic errors $u_{x,i,t}$ and $u_{w,i,t}$ are $p_x\times 1$ and $p_w\times 1$, respectively. The presence of $f_t$ here is reasonable because regressors are often co-moving, both among themselves and across the cross-section.

A prominent example of an unobserved factor that affects both bank lending and the right-hand side control variables is the stimulus provided by the Federal Reserve through conventional monetary policy, as measured by the difference between the federal funds rate and the optimal interest rate predicted by models such as the Taylor rule. This variable is unobserved and is therefore typically not included. Other factors include concurrent policies such as the Troubled Asset Relief Program, house price growth, and aggregate demand and supply shocks from current and anticipated macroeconomic conditions. It is natural to assume that banks respond heterogeneously to each of these factors, which in \eqref{eq:e} is captured by the unit-specific factor loadings $\gamma_i$, $\Gamma_{w,i}$ and $\Gamma_{x,i}$. This is a departure from the popular but restrictive two-way error components model which clumps shocks to a single factor that has an equal effect on all units.

The formal assumptions that we will be working under are laid out in Section \ref{section::assump}. However, before we take the assumptions it is useful to first stack \eqref{eq:y}--\eqref{eq:w} over time, and to also introduce the CCE estimator that we will be using. Let us therefore denote by $\mathrm{diag}(A, B)$ the block-diagonal matrix that takes the matrices $A$ and $B$ as the upper left and lower right block, respectively. Let $\mathcal{T}_k=\{T_1,...,T_k\}$ be the set of breakpoints. We can now define the following matrices:
\begin{align*}
	W_i(\mathcal{T}_k) & = \mathrm{diag}(W_{1,i} ,..., W_{k+1,i}), ~~ U_{w,i}(\mathcal{T}_k)= \mathrm{diag}(U_{w,1,i} ,..., U_{w,k+1,i}),\\
	F(\mathcal{T}_k) &= \mathrm{diag}(F_{1} ,..., F_{k+1}),
\end{align*}
where $W_{j,i}=(w_{i,T_{j-1}+1},...,w_{i,T_j})'$ is $(T_j-T_{j-1})\times p_w$, $U_{w,j,i}=(u_{w,i,T_{j-1}+1},...,u_{w,i,T_j})'$ is $(T_j-T_{j-1})\times p_w$, and $F_{j}=(f_{T_{j-1}+1},...,f_{T_j})'$ is $(T_j-T_{j-1})\times m$ for $j=1,...,k+1$. Hence,
$W_{j,i}$, $U_{w,j,i}$ and $F_{j}$ stacks the time observations within regime $j$. The matrices $W_i(\mathcal{T}_k)$, $U_{w,i}(\mathcal{T}_k)$ and $F(\mathcal{T}_k)$ stack these regime-specific stacks on the diagonal, which means that their dimensions must be $T\times (k+1)p_w$, $T\times (k+1)p_w$ and $T\times (k+1)m$, respectively. We also introduce the $T\times 1$ vectors $Y_i=(y_{i,1},...,y_{i,T})' $ and $ e_i=(e_{i,1},...,e_{i,T})' $, the $T\times p_x$ matrix $X_i=(x_{i,1}',...,x_{i,T}')'$, the $(k+1)p_w \times 1$ vector $\delta=(\delta_1',...,\delta_{k+1}')'$, the $T\times m$ matrix $F=(f_1,...,f_T)'$, and the $T\times p_x$ matrix $U_{x,i}=(u_{x,i,1},...,u_{x,i,T})'$. In this notation, the time-stacked version of \eqref{eq:y}--\eqref{eq:w} becomes
\begin{align}
	Y_i&=X_i\beta+W_i(\mathcal{T}_k)\delta+e_i, \label{eq:yst}\\
	e_i&=F\gamma_i+\varepsilon_i,  \label{eq:est}\\
	X_i&=F \Gamma_{x,i}+U_{x,i}, \label{eq:xst}\\
	W_i(\mathcal{T}_{k})&=F(\mathcal{T}_{k})(I_{k+1} \otimes \Gamma_{w,i})+U_{w,i}(\mathcal{T}_{k}). \label{eq:wst}
\end{align}

The fact that $F$ enters \eqref{eq:est}--\eqref{eq:wst} means that \eqref{eq:yst} cannot be consistently estimated by ordinary least squares (OLS) even if $\mathcal{T}_{k}$ is known, because the $X_{i}$ and $W_i(\mathcal{T}_{k})$ are correlated with $e_{i}$ through $F$. Consistent estimation of $\beta$ and $\delta$ is therefore not possible without properly controlling for $F$. Two of the most popular approaches that can be used for this purpose are the principal components-based approach of Bai (2009) and the CCE approach of Pesaran (2006). In the present paper, we opt for the latter approach, because of its simplicity, robustness and excellent small-sample properties (see, for example, Westerlund and Urbain, 2015).

The original CCE estimator uses cross-sectional averages of the dependent variable and the regressors to estimate the (space spanned by the) unknown factors. As Karavias et al. (2022) point out, however, this is not the best approach when there are breaks present, as the cross-sectional average of the dependent variable is uninformative about $F$. The CCE estimator that we use here is therefore based on the cross-section averages of the regressors only. Let us collect these regressors in the $T\times (p_x+(k+1)p_w)$ matrix $Z_i (\mathcal{T}_k)=(X_i,W_i(\mathcal{T}_k))$. In view of \eqref{eq:xst} and \eqref{eq:wst}, it is not difficult to see that the data generating process of $Z_i(\mathcal{T}_{k})$ is given by
\begin{equation}
	Z_i(\mathcal{T}_{k})=F_Z(\mathcal{T}_{k}) \Gamma_i+ U_{i}(\mathcal{T}_{k}),
	\label{eq:factorrep}
\end{equation}
where $F_Z (\mathcal{T}_{k})= (F,F(\mathcal{T}_{k}) )$ is $T\times (k+2)m$, $\Gamma_i=\Gamma_i(\mathcal{T}_{k})=\mathrm{diag}(\Gamma_{x,i} , I_{k+1} \otimes \Gamma_{w,i})$ is $(k+2)m\times (p_x+(k+1)p_w)$ and $U_i(\mathcal{T}_k)=(U_{x,i},U_{w,i}(\mathcal{T}_{k}))$ is $T\times (p_x+(k+1)p_w)$. Denote by $\bar A = N^{-1}\sum_{i=1}^{N}A_i$ the cross-sectional average of any generic variable $A_i$. By a law of large numbers, $\bar U(\mathcal{T}_{k})$ should be ``close'' to zero under standard conditions, which by \eqref{eq:factorrep} means that $\bar Z (\mathcal{T}_k)$ should be close to $F_Z(\mathcal{T}_{k})\bar\Gamma$. Define the following $T\times T$ the projection error matrix:
\begin{equation}
	M_{\bar Z (\mathcal{T}_{k})}=I_T-\bar Z(\mathcal{T}_k) (\bar Z(\mathcal{T}_k)'\bar Z(\mathcal{T}_k))^{-1} \bar Z(\mathcal{T}_k)'.
\end{equation}
The above discussion suggests that pre-multiplication by this matrix should be enough to remove the factors, at least asymptotically. Let us therefore denote by $\tilde{A} = \tilde{A}(\mathcal{T}_k) = M_{\bar Z(\mathcal{T}_k)} A$ the ``defactored'' version of any $T$-rowed matrix $A$.\footnote{The dependence of $\tilde{A}$ on $\mathcal{T}_k$ is suppressed for notational simplicity.} In this notation, the model to be estimated is given by
\begin{align}
\tilde Y_i =\tilde X_i\beta+\tilde W_i(\mathcal{T}_k)\delta + \tilde e_i.
	\label{eq:stackedt_tilde}
\end{align}
This model can be stacked over the cross-section, giving
\begin{align}
	\tilde Y =\tilde X\beta+\tilde W(\mathcal{T}_k)\delta+\tilde E,
	\label{eq:stackedt_tilde_full}
\end{align}
where $\tilde Y = (\tilde Y_1',...,\tilde Y_N')'$, $\tilde X = (\tilde X_1',...,\tilde X_N')'$, $\tilde W(\mathcal{T}_k) = (\tilde W_1(\mathcal{T}_k)',...,\tilde W_N(\mathcal{T}_k)')'$ and $\tilde E = (\tilde e_1',...,\tilde e_N')'$ are all $NT$-rowed. The CCE estimator of $\delta$ that we will be considering in this paper is simply the pooled OLS estimator obtained from \eqref{eq:stackedt_tilde_full};
\begin{equation}
	\hat{\delta}(\mathcal{T}_k)= (\tilde W(\mathcal{T}_k)'M_{\tilde X} \tilde W(\mathcal{T}_k) )^{-1} \tilde W(\mathcal{T}_k)' M_{\tilde X} \tilde Y, 	\label{eq:deltahat}
\end{equation}
where $M_{\tilde X}=I_{NT}-\tilde X(\tilde X'\tilde X)^{-1}\tilde X'$.

The estimated factors in $\bar Z (\mathcal{T}_k)$ depends on the unknown breakpoints, as does $W(\mathcal{T}_k)$. As we show later in Section \ref{section::breakestim}, however, the estimated breakpoints are consistent, which means that asymptotically $\mathcal{T}_k$ can be treated as known. The factors can therefore be treated as known, too. This is the reason for why the proposed CCE approach works even if in practice both the breaks and the factors are unknown.

\subsection{Assumptions}\label{section::assump}

We begin this section with some additional notation. If $A$ is a matrix, $\mathrm{tr}( A)$ and $\mathrm{rank}(A)$ signify its trace and rank, respectively, and $\|A\| = \sqrt{\mathrm{tr} (A'A)}$ signifies its Frobenius norm. We write $A > 0$ to signify that $A$ is positive definite. The symbols $\to_d$, $\to_p$ and $\to_w$ signify convergence in distribution, convergence in probability and weak convergence, respectively, and $\lfloor x\rfloor$ signifies the integer part of $x$. We use $N,\, T \to\infty$ to indicate that the limit has been taken while passing both $N$ and $T$ to infinity. We use w.p.a.1 to denote with probability approaching one. Here and throughout the reminder of this paper, $\mathcal{T}^0_{k^0}=\{T_1^0,...,T_{k^0}^0\}$ will be used to denote the set of true breakpoints with $k^0$ being the true number of breaks. If $k^0  =0$, we define $\mathcal{T}^0_{k^0}= \emptyset$.

\begin{assumption}[Breaks]\
	\begin{enumerate}[(i)]
\item $T_j^0=\lfloor \lambda_j^0 T \rfloor$ for $j=1,...,k^0+1$, where $\lambda_0^0 = 0 < \lambda_1^0 < ... <\lambda_{k}^0 < \lambda_{k^0+1}^0 = 1$.

\item $0 < \|\delta\| < \infty$ and $\delta_{j+1} \ne \delta_j$ whenever $k^0 > 0$.
	\end{enumerate}\label{ass:breakdates}
\end{assumption}

\begin{assumption}[Errors]\
	\begin{enumerate}[(i)]
		\item $u_{i,t}=(u_{x,i,t}',u_{w,i,t}')'$ is a covariance stationary process that is independent across $i$ with absolutely summable autocovariances, $E(u_{i,t}) = 0_{(p_x+p_w)\times 1}$, $E(u_{i,t}u_{i,t}')=\Sigma_{u,i}$ and $E(\|u_{i,t}\|^4)< \infty$.
		
		\item $\varepsilon_{i,t}$ is a covariance stationary process that is independent across $i$ with absolutely summable autocovariances, $E(\varepsilon_{i,t})=0$, $E( \varepsilon_{i,t}^2 ) =\sigma_{\varepsilon,i}^{2}$, $E(\varepsilon_i \varepsilon_i')=\Sigma_{\varepsilon,i}$ and $E(\varepsilon_{i,t}^4) < \infty$.
		
		\item $\varepsilon_{i,t}$ and $u_{j,s}$ are independent for all $i$, $j$, $s$ and $t$.
	\end{enumerate}\label{ass:err}
\end{assumption}

\begin{assumption}[Factors]\
	\begin{enumerate}[(i)]
		\item $T^{-1}F'F > 0$ w.p.a.1 for all $T$.

		\item $E(\|f_t\|^2)<\infty$ for all $t$.

		\item $f_t$ is independent of $\varepsilon_{i,s}$ and $u_{i,s}$ for all $i$, $s$ and $t$.
	\end{enumerate}\label{ass:fact}
\end{assumption}

\begin{assumption}[Loadings]\
	\begin{enumerate}[(i)]
		\item $\mathrm{rank}(\bar \Gamma_x) = m\leq p_x$ and $\mathrm{rank}(\bar \Gamma_w)=m\leq p_w$ for all $N$.
		
		\item $\gamma_{i}=\gamma+\eta_{i}$ and $\Gamma_{i}=\Gamma+\xi_{i}$, where $\eta_{i}$ and $\xi_{i}$ are independently distributed across $i$ with $E(\eta_i) = 0_{m\times 1}$, $E(\xi_i) = 0_{(k^0+2)m\times (p_x+(k^0+1)p_w)}$, $E(\|\eta_i\|^2) < \infty$ and $E(\|\xi_i\|^2) < \infty$.
		
		\item $\gamma_{i}$ and $\Gamma_{i}$ are independent of $\varepsilon_{j,t}$, $u_{j,t}$ and $f_t$ for all $i$, $j$ and $t$.
	\end{enumerate}\label{ass:load}
\end{assumption}

\begin{assumption}[Invertibility]\
$(NT)^{-1} \tilde X'\tilde X > 0$ and $(NT)^{-1} \tilde W(\mathcal{T}_k)' M_{\tilde X}\tilde W(\mathcal{T}_k) > 0$ w.p.a.1 for all $N$ and $T$.\label{ass:invert}
\end{assumption}

\begin{assumption}[Moments]\ \label{ass:limit}
The following holds as $N,\,T\to\infty$:
	\begin{enumerate}[(i)]
  \item $\begin{aligned}[t]
  \frac{1}{N\Delta T_j^0}\sum_{i=1}^N \sum_{t=T_{j-1}^0+1}^{T_{j}^0}\check u_{w,i,t}\check u_{w,i,t}' \to_p \Omega_j, \\
    \end{aligned}$

  \item $\begin{aligned}[t]
\frac{1}{N\Delta T_j^0}\sum_{i=1}^N \sum_{t=T_{j-1}^0+1}^{T_{j}^0}\sum_{l=T_{j-1}^0+1}^{T_{j}^0} \varepsilon_{i,t}\varepsilon_{i,l}\check u_{w,i,t}\check u_{w,i,l}' \to_p \Phi_j ,  \end{aligned}$

  \item $\begin{aligned}[t]
\frac{1}{\sqrt{N\Delta T_j^0}}\sum_{i=1}^N \sum_{t=T_{j-1}^0+1}^{T_{j-1}^0+ \lfloor s\Delta T_j^0\rfloor} \varepsilon_{i,t}\check u_{w,i,t} \to_w \Phi_j^{1/2}B_j(s), \end{aligned}$
	\end{enumerate}
where $j=1,...,k^0+1$, $\Delta T_{j}^0= T_{j}^0-T_{j-1}^0$, $s\in [0,1]$, $B_i(s)$ is an $p_w\times 1$ vector standard Brownian motion on the same interval, and $\check u_{w,i,t} = u_{w,i,t} - \sum_{n=1}^N \sum_{l=T_{j-1}^0+1}^{T_{j}^0} u_{w,n,l}a_{i,n,t,l}$ with $a_{i,n,t,l} = u_{x,n,l}'(\sum_{i=1}^N U_{x,i}'U_{x,i})^{-1}u_{x,i,t}$.
\end{assumption}

Assumption \ref{ass:breakdates} requires that the breaks are distinct and hence that each regime increases with $T$. This is standard in the type of large-$T$ panel data that we are considering (see, for example, Baltagi et al., 2016). In pure time series, it is even necessary, as enough observations are needed to consistently estimate the slope coefficients in each regime. The panel data structure brings more (cross-sectional) information to the table and this enables consistent estimation of the breakpoints even if the regimes are not expanding (as shown in the online appendix). Testing for breaks is, however, more demanding in this regard and is not possible without Assumption \ref{ass:breakdates}.

Assumptions \ref{ass:err}--\ref{ass:load} are standard in the CCE literature (see Pesaran, 2006). Some of these can be relaxed at the expense of more complicated proofs but not all. For example, Assumption \ref{ass:fact} (i) and (ii) can be relaxed to allow for more general types of (non-stationary) factors (see Westerlund, 2018). Assumption \ref{ass:load} (i) is also stronger than necessary. As we show in the online appendix, it implies that $\mathrm{rank}(\bar \Gamma)=(k^0+2)m\leq p_x+(k^0+1)p_w$, which means that the number of factors appearing in \eqref{eq:factorrep} cannot be larger than the number of averages in $\bar Z(\mathcal{T}_{k})$ employed by CCE to estimate those factors. This condition can be relaxed if some of the factors are observed. Observed factors can be appended to $\bar Z(\mathcal{T}_{k})$ and projected out with the effect that these factors do not have to satisfy Assumption \ref{ass:load} (i). In Section \ref{section::appl} we elaborate on this point.

Assumption \ref{ass:invert} is a non-collinearity condition. It demands that the regressors have enough variation across both the cross-section and time after projecting out all variation that can be explained by the factors, which is a standard requirement in the interactive effects literature (see, for example, Bai, 2009, and Pesaran, 2006). This generalizes the within variation assumption in the conventional one- and two-way error component models.

Assumption \ref{ass:limit} requires that the moments of $u_{w,i,t}$ after projecting out the part of the variation that is due to $u_{x,i,t}$ is constant within regimes. This is needed for our asymptotic distribution theory, but not for the consistency of the estimated breakpoints.

\section{The toolbox and its asymptotic properties}\label{section::toolbox}

\subsection{Testing for structural breaks}\label{section::hypotheses}

This section presents tests for three hypotheses, labelled ``(A)''--``(C)'', which are useful for establishing whether or not there are any breaks and for determining their number. Some of these are stated in terms of the set of permissible break dates, in which all breaks are distinct and bounded away from the sample endpoints. This set is given by
\begin{equation}
\mathcal{T}_{k,\epsilon} = \{(T_1,...,T_k):\Delta T_{j+1}  \geq \epsilon T, T_1\geq \epsilon T, T_k\leq (1-\epsilon) T \},
\end{equation}
where $\Delta T_{j}= T_{j}-T_{j-1}$ and $\epsilon > 0$ is a user-defined trimming parameter, the choice of which will be discussed later.

\bigskip

\noindent \textbf{Hypotheses:}

\begin{enumerate}[(A)]
\item The null hypothesis is that there are no breaks and the alternative hypothesis is that there are at most $k$ structural breaks, which may or may not be equal to $k^0$, and where the dates of the breaks can be either known or unknown. If the breakpoints are known, the null and alternative hypotheses can be stated in the following way:
\begin{align}
H_0^A &:  \delta_1=...=\delta_{k+1} ,\\
H_1^A &:  \delta_n \neq \delta_j \ \text{for some}\ n\neq j \in \{1,...,k+1\}.
\end{align}
If, on the other hand, the breakpoints are unknown, the null hypothesis is the same but the alternative changes to
\begin{align}
	H_1^A: \bigcup_{\mathcal{T}_k\in \mathcal{T}_{k,\epsilon}}\{\delta_n \neq \delta_j \ \text{for some}\ n\neq j \}.
\end{align}

\item The null hypothesis is that there are no breaks and the alternative is that there is an unknown number of breaks, where the number of breaks is bounded from above by some prescribed value $k_{max}$. Formally,
\begin{align}
H_0^B &: \bigcup_{k=2}^{k_{max}+1}\{\delta_1=...=\delta_{k} \},\\
H_1^B &: \bigcup_{k=1}^{k_{max}}\left\{\bigcup_{\mathcal{T}_k\in \mathcal{T}_{k,\epsilon}}\delta_n \neq \delta_j \ \text{for some}\ n\neq j\right\}.
\end{align}

\item The null hypothesis is that there are $k$ breaks and the alternative is that there are $k+1$ breaks. The value of $k$ is specified by the researcher. Formally,
\begin{align}
H_0^C &:\bigcup_{i=1}^{k+1}\left\{\delta_i=\delta_{i+1} \ \text{and} \ \delta_j\neq \delta_k \ \text{for any} \ (j,k)\neq(i,i+1) \right\},\\
H_1^C &: \bigcap_{\substack{i,j\in \{1,...,k+1\}\\i\neq j}}\left\{\delta_i\neq\delta_j \right\}.
\end{align}
\end{enumerate}

We will consider four test statistics, two for (A) and one for each of (B) and (C). These can be seen as panel extensions of the time series statistics developed by BP98.

To test $H_0^A$ versus $H_1^A$ when the dates of the breaks are known, define the $kp_w\times(k+1)p_w$ matrix $R= I_k \otimes ( I_{p_w} , -I_{p_w} )$, which is such that $(R\delta)' = (\delta_1'-\delta_2', ..., \delta_k'-\delta_{k+1}')$. The appropriate $F$-statistic to use for testing this hypothesis is given by $F(\mathcal{T}_{k^0}^0)$, where
\begin{equation}
F(\mathcal{T}_k)=\frac{N(T-p_x-(k+1)p_w)-p_x-(k+1)p_w}{kp_w}  \hat \delta(\mathcal{T}_k)'R'(R \hat{V}_{\hat\delta} R')^{-1}R\hat\delta(\mathcal{T}_k).
\label{eq:wald}
\end{equation}
The degree of freedom correction used in the normalization of $\hat \delta(\mathcal{T}_k)'R'(R \hat{V}_{\hat\delta} R')^{-1}R\hat\delta(\mathcal{T}_k)$ is not necessary but we keep it since it leads to slightly better small sample performance than if $NT/(kp_w)$ is used. The $(k+1)p_w \times (k+1)p_w$ matrix $\hat{V}_{\hat\delta}$ is an estimator of the asymptotic covariance matrix of $\hat \delta(\mathcal{T}_k)$ (with the dependence on $\mathcal{T}_k$ suppressed), which is given by
\begin{equation}
\hat{V}_{\hat\delta} = \hat \Omega^{-1} \hat{\Phi} \hat \Omega^{-1}.
\end{equation}
Here
\begin{align}
\hat \Omega & = (NT)^{-1}\tilde{W}(\mathcal{T}_k)'M_{\tilde{X}}\tilde{W}(\mathcal{T}_k),\\
\hat{\Phi} & = \hat{\Lambda}_{0}+\sum_{l=1}^{L}\left(1-\frac{l}{L+1}\right)(\hat \Lambda_{l}+\hat \Lambda_{l}'), \label{eq:nwbreak}\\
\hat{\Lambda}_{l} & = \frac{1}{NT}\sum_{i=1}^{N}\sum_{t=l+1}^{T}\hat{\varepsilon}_{i,t}\hat{\varepsilon}_{i,t-l} \check w_{i,t}\check w_{i,t-l}',
\label{eq:nwbreak1}
\end{align}
where $L$ is a user-specified bandwidth. The $1\times p_w$ vector $\check{w}_{i,t}'$ is the $t$-th row of the $T\times p_w$ matrix $\check{W}_{i}=(\check{w}_{i,1},...,\check{w}_{i,T})'$, which in turn is the $i$-th block of the $NT\times p_w$ matrix $M_{\tilde{X}}\tilde{W}(\mathcal{T}_k) = (\check{W}_{1}',...,\check{W}_{N}')'$. The scalar $\hat{\varepsilon}_{i,t}$ is the $t$-th row of the $T\times 1$ vector $\hat \varepsilon_i=(\hat{\varepsilon}_{i,1},,...,\hat{\varepsilon}_{i,T})'$, which is the $i$-th block of the $NT\times 1$ vector $\hat \varepsilon=(\hat \varepsilon_1',...,\hat \varepsilon_{N}')'=M_{\tilde{X}}(\tilde Y-\tilde{W}(\mathcal{T}_k)\hat\delta(\mathcal{T}_k))$.

The following theorem provides the asymptotic distribution of $F(\mathcal{T}_k^0)$, which is in turn key in deriving the asymptotic distributions of the other tests that we will be considering. The theorem is stated in terms of the following process, which is a multiple break generalization of the squared tied-down Bessel process that is otherwise so common in the break testing literature (see, for example, Andrews, 1993):
\begin{equation}
Q(\Lambda_{k})=\frac{1}{kp_w}\sum_{j=1}^{k}\frac{ [\lambda_j B(\lambda_{j+1})-\lambda_{j+1} B(\lambda_{j})]'[\lambda_j B(\lambda_{j+1})-\lambda_{j+1} B(\lambda_{j})]}{\lambda_j\lambda_{j+1}(\lambda_{j+1}-\lambda_j)},
\end{equation}
with $B(\lambda_j)$ being an $p_w\times 1$ vector standard Brownian motion on $[0,1]$ and  $\Lambda_{k}=\{\lambda_1,...,\lambda_{k}\}$. The true value of $\Lambda_{k}$ is denoted $\Lambda_{k^0}^0=\{\lambda_1^0,...,\lambda_{k^0}^0\}$. We also define the supremum of the above process;
\begin{equation}
\mathrm{sup} Q(k) = \sup_{\Lambda_k\in \Lambda_{k,\epsilon}} Q(\Lambda_k),
\end{equation}
where
\begin{equation}
\Lambda_{k,\epsilon} = \{(\lambda_1,...,\lambda_k): \Delta\lambda_{j+1}  \geq \epsilon , \lambda_1\geq \epsilon, \lambda_k\leq 1-\epsilon  \}
\end{equation}
with $ \Delta\lambda_{j} = \lambda_{j} - \lambda_{j-1}$. The dependence of $Q(\Lambda_{k})$ and $\mathrm{sup} Q(k)$ on $p_w$ and $\epsilon$ is suppressed in order to avoid cluttering the notation.

\begin{theorem}\label{thm:f}
Suppose that Assumptions \ref{ass:breakdates}--\ref{ass:limit} are met, and that $H_0^A$ holds. Then, as $N,\,T\to\infty$ with $T/N \to 0$,
\begin{align}
F(\mathcal{T}_{k^0}^0) \to_w Q(\Lambda_{k^0}^0) =_d F_{k^0p_w,N(T-p_x-(k^0+1)p_w)-p_x-(k^0+1)p_w},
\end{align}
where $\to_w$ and $=_d$ signify weak convergence and equality in distribution, respectively.
\end{theorem}

We require that $T/N \to 0$, as otherwise the error coming from the estimation of the factors will tend to accumulate as we sum over time. It is therefore necessary. In practice, however, having $N>> T$ do not seem to be very important. Indeed, as we demonstrate in our Monte Carlo study reported in the online appendix, the proposed toolbox seems to perform well even when $ N $ and $ T $ are similar in size, which is consistent with existing evidence for the CCE approach (see, for example, Pesaran, 2006, and Westerlund and Urbain, 2015). Moreover, many data sets like the one used in Section \ref{section::appl} do in fact have $ N $ larger than $ T $. The condition that $T/N \to 0$ is therefore unlikely to pose a problem in empirical work. 

If $\mathcal{T}_k^{0}$ is unknown, which is the empirically most plausible scenario, the following statistic may be used:
\begin{equation}
\mathrm{sup}F(k)= \sup_{\mathcal{T}_k\in \mathcal{T}_{k,\epsilon}} F(\mathcal{T}_k).
\label{eq:supwald}
\end{equation}
This test is feasible if $k$ is ``small''. If $k$ is ``large'' it becomes computationally very costly to find the set $\mathcal{T}_k$ that maximises $F(\mathcal{T}_k)$. Grid search requires $O(T^k)$ least squares operations. By contrast, by using the efficient breakpoint estimation algorithm presented in the next subsection, Section \ref{section::breakestim}, we can limit the number of operations to $O(T^2)$ for any $k$. The basic idea is to first apply this algorithm to obtain $\hat{\mathcal{T}}_k=\{\hat T_1,...,\hat T_{k}\}$. Analogously to $T_0$, $T_{k+1}$ and $\mathcal{T}_0$, we define $\hat T_0=1$, $\hat T_{k+1}=T$ and $\hat{\mathcal{T}}_0 = \emptyset$. Given $\hat{\mathcal{T}}_k$, we compute $\mathrm{sup}F(k) = F(\hat{\mathcal{T}}_k)$ and use this as our test statistic. The asymptotic distribution of $\mathrm{sup}F(k)$ is a direct consequence of Theorem \ref{thm:f} and it is reported in Corollary \ref{cor:dist} below. However, before we state the corollary, we present the proposed tests of hypotheses (B) and (C).

Testing $H_0^B$ versus $H_1^B$ can be done using the following panel version of BP98's ``weighted double maximum'' statistic:
\begin{equation}
\mathrm{WDmax}F(k_{max})= \max_{1\leq k\leq k_{max}} \frac{c_{\alpha,1}}{c_{\alpha,k}} \mathrm{sup}F(k),
\label{eq:doublemax}
\end{equation}
where $c_{\alpha,k}$ is the critical value of $\mathrm{sup}F(k)$ at significance level $\alpha$ and $k$ breaks. The weighting by $c_{\alpha,1}/c_{\alpha,k}$ here ensures that the marginal $p$-values of the weighted supremum statistics are all equal. This counterweights the decrease in the marginal $p$-value of $\mathrm{sup}F(k)$ that comes from increasing $k$, and the resulting loss of power when $k$ is large.

The test of $H_0^C$ versus $H_1^C$ can be carried out using the following statistic:
\begin{equation}
F(k+1|k)= \sup_{1\leq j \leq k+1} \sup_{T_+ \in \hat{\mathcal{T}}_{j,\epsilon}} F(T_+|\hat{\mathcal{T}}_k).
\label{eq:seqf}
\end{equation}
where $\hat{\mathcal{T}}_{k}$ is a set of $k$ estimated (or known) break dates stipulated under the hull hypothesis, $T_+$ is the additional $(k+1)$-th break under the alternative, and
\begin{equation}
\hat{\mathcal{T}}_{j,\epsilon} = \{T_+ : \hat T_{j-1}+ \Delta\hat T_j \epsilon \leq T_+ \leq  \hat T_{j}- \Delta\hat T_j\epsilon \}
\end{equation}
is the set of permissible breaks in between the estimated $(j-1)$-th and $j$-th breaks. Hence, $F(k+1|k)$ is testing the null of $k$ breaks versus the alternative that there is an additional break somewhere within the regimes stipulated under the null. Finally, $F(T_+|\hat{\mathcal{T}}_k)$ is given by
\begin{align}
F(T_+|\hat{\mathcal{T}}_k) & =\frac{N(T-p_x-(k+2)p_w)-p_x-(k+2)p_w}{p_w} \notag\\
& \times \hat \delta(\{\hat{\mathcal{T}}_k,T_+\})'R_j'(R_j \hat{V}_{\hat \delta} R_j')^{-1}R_j\hat\delta(\{\hat{\mathcal{T}}_k,T_+\}),
\end{align}
where $\hat\delta(\{\hat{\mathcal{T}}_k,T_+\})$ comes from a regression with $k+1$ breaks at dates $\{\hat{\mathcal{T}}_k,T_+\}$. The same is true for $\hat{V}_{\hat \delta}$. The matrix $R_j$ is given by $R_j=(0_{p_w\times 1}, ..., 0_{p_w\times 1}, I_{p_w}, -I_{p_w}, 0_{p_w\times 1}, ..., 0_{p_w\times 1})$, where $I_{p_w}$ sits in the $j$-th position, while $-I_{p_w}$ sits in the $(j+1)$-th, so that $(R_j\delta)' = \delta_j'-\delta_{j+1}'$.\footnote{If the errors are homoskedastic and serially uncorrelated, then $F(k+1|k)$ takes the same form as in equation (10) in BP98.}

\begin{corollary}\label{cor:dist}
Suppose that Assumptions \ref{ass:breakdates}--\ref{ass:limit} are met, and that the null hypothesis of each test holds. Then, as $N,\,T\to\infty$ with $T/N \to 0$,
\begin{itemize}
  \item[(a)] $\begin{aligned}[t]
  \mathrm{sup}F(k) \to_w \mathrm{sup} Q(k), \\
    \end{aligned}$

  \item[(b)] $\begin{aligned}[t]
\mathrm{WDmax}F(k_{max}) \to_w \max_{1\leq k\leq k_{max}} \frac{c_{\alpha,1}}{c_{\alpha,k}} \mathrm{sup} Q(k),     \end{aligned}$

  \item[(c)] $\begin{aligned}[t]
P[ F(k+1|k) \leq x ]  \to (P[\mathrm{sup} Q(1)\leq x ])^{k+1}.  \end{aligned}$
\end{itemize}
\end{corollary}

The results in (a) and (b) follow from Theorem \ref{thm:f} and the continuous mapping theorem. The intuition behind (c) goes as follows: The $F(k+1|k)$ test amounts to $k+1$ tests of the null hypothesis of no break versus the alternative of a singe break, each of which converges to $\mathrm{sup} Q(1)$ under the null. Moreover, because the sample segments on which the tests are applied are non-overlapping, the asymptotic distributions are independent. The stated result for $P[ F(k+1|k) \leq x ]$ then follows from standard results for order statistics.

The asymptotic results given in Corollary \ref{cor:dist} are the same as the ones given in Propositions 6 and 7 of BP98. This is convenient because it means that appropriate critical values are already available. Critical values for $\mathrm{sup}F(k)$ and $\mathrm{WDmax}F(k_{max})$ are reported for $\epsilon = 0.05$, $k\in\{1,...,9\}$ and $p_w\in\{1,...,10\}$ in Table I of BP98, and in Table II they report critical values for $F(k+1|k)$. Bai and Perron (2003a, Table 1) report response surface regressions for all tests that are valid for more values $\epsilon$, $k$ and $p_w$.

The $F(k+1|k)$ test can be applied sequentially for $k=0,1,...$ to estimate the number of breaks. In this case, we start by testing the null of no breaks against the alternative of a single break using $F(1|0)$. If the null is accepted, we set $\hat k = 0$ and terminate the procedure. If, however, the null is rejected, we estimate the breakpoint, denoted $\hat T_1$, and split the sample in two at $\hat T_1$. We then test for the presence of a break in each of the two subsamples using $F(2|1)$. If no breaks are found, we set $\hat k = 1$ and stop, whereas if breaks are detected, we estimate their location and split the sample again. This process continues until the test fails to reject, or until the maximum permissible number of breaks is reached. This number is a function of the trimming parameter $ \epsilon $ and is given by $ \left\lfloor 1/\epsilon \right\rfloor-2 $, where $ \left\lfloor 1/x \right\rfloor $ denotes the integer part of $ x $.

A problem with the sequential approach just described is that it does not account for the multiplicity of the testing problem. It will therefore reject too often. In order to prevent this from happening the significance level of each test in the sequence, $\alpha$ say, should be set as a decreasing function of the sample size.

\begin{theorem}\label{thm:khat}
Suppose that Assumptions \ref{ass:breakdates}--\ref{ass:limit} are met. Suppose also that $F(k+1|k)$ is applied sequentially at significance level $\alpha$ such that $NT \alpha \to K > 0$. Then, as $N,\,T\to\infty$ with $T/N \to 0$,
\begin{equation}
P(\hat k = k^0) \to 1.
\end{equation}
\end{theorem}

Theorem \ref{thm:khat} requires that $\alpha$ converges to zero slowly enough. Of course, in practice $N$ and $T$ are always fixed, and hence so is $\alpha$. Our simulation results show that $\alpha = 0.05$ works satisfactorily, as do the results of Bai (1999).

\subsection{Breakpoint estimation}\label{section::breakestim}

In the previous section, we were concerned with testing for the existence of breaks, and with estimating their number, $k^0$. Once $k^0$ has been estimated, however, interest turns to the location of the breaks, and this is the topic of the present section. According to Theorem \ref{thm:khat}, knowing $\hat k$ is as good as knowing $k^0$, at least asymptotically. In this section, we therefore treat $k^0$ as known.

The problem of estimating the breakpoints is not independent of that of testing for breaks. In fact, the $\hat{\mathcal{T}}_k$ that minimizes the sum of squared residuals (for a given $k$) is the same as the one that maximizes $F(\mathcal{T}_k)$. The breakpoint estimator that we will employ is therefore given by
\begin{equation}
\hat{\mathcal{T}}_{k^0} =\arg \min_{\mathcal{T}_{k^0} \in \mathcal{T}_{k^0,\epsilon}} SSR(\mathcal{T}_{k^0}) =\arg \max_{\mathcal{T}_{k^0} \in \mathcal{T}_{k^0,\epsilon}} F(\mathcal{T}_{k^0}), \label{eq:breakdateestim}
\end{equation}
where 
\begin{equation}
SSR(\mathcal{T}_k)=\hat \varepsilon'\hat \varepsilon = (\tilde Y-\tilde{W}(\mathcal{T}_k)\hat\delta(\mathcal{T}_k))'M_{\tilde{X}}(\tilde Y-\tilde{W}(\mathcal{T}_k)\hat\delta(\mathcal{T}_k)).
\end{equation}

Consider the model for $\tilde Y$ in \eqref{eq:stackedt_tilde_full}. If $\beta = 0_{p_x\times 1}$ was known, then $\tilde Y=\tilde W(\mathcal{T}_{k^0})\delta+\tilde E$ is a pure structural change model, and for such a model $\mathcal{T}_{k^0}$ and $\delta$ can be estimated using the dynamic programming algorithm of Bai and Perron (2003b), ``BP03'' henceforth. While initially proposed as an efficient way to minimize the sum of squared residuals in the pure time series context, the algorithm can be extended in a straightforward manner to the current more general context, as it is just a way to compare possible combinations of breakpoints to achieve a minimum global sum of squared residuals. The efficiency of the algorithm comes from recognizing that with $T$ times series observations the total number of possible sample splits is $T(T+1)/2$ for any $k$ and is therefore $O(T^2)$. With $\beta$ known but not necessarily zero, then $\tilde X\beta$ can be subtracted from $\tilde Y$ and the dynamic programming algorithm can be applied to $\tilde Y -\tilde X\beta =\tilde W(\mathcal{T}_{k^0})\delta+\tilde E$, which is again a pure structural change model. With $\beta$ unknown, the estimation can be carried out in an iterative fashion, which we will now describe.

\bigskip

\noindent \textbf{Breakpoint estimation algorithm:}

\begin{enumerate}
\item Initiate $\hat \beta$ by treating the coefficients of both $x_{i,t}$ and $w_{i,t}$ as subject to structural change. Define $\tilde X(\mathcal{T}_{k^0})$ similarly to $\tilde W(\mathcal{T}_{k^0})$, and apply BP03's dynamic programming algorithm to the pure structural change model $\tilde Y= \tilde X(\mathcal{T}_{k^0}) \beta + \tilde W(\mathcal{T}_{k^0})\delta + \text{error}$, where $(\bar X(\mathcal{T}_{k^0}), \bar W(\mathcal{T}_{k^0}))$ is used in place of $\bar Z(\mathcal{T}_{k^0})$ to estimate the factors. This yields $\hat{\mathcal{T}}_{k^0}$, $\hat \beta$ and $\hat \delta$.

\item Update $\hat \beta$ by fitting $\tilde Y -\tilde W(\hat{\mathcal{T}}_{k^0})\hat{\delta} = \tilde X\beta + \text{error}$ by OLS, this time using $\bar X$ in place of $\bar Z(\mathcal{T}_{k^0})$.

\item Update $\hat{\mathcal{T}}_{k^0}$ and $\hat \delta$ by applying the dynamic programming algorithm to $\tilde Y -\tilde X\hat{\beta} = \tilde W(\mathcal{T}_{k^0})\delta + \text{error}$ using $\bar W(\mathcal{T}_{k^0})$ in place of $\bar Z(\mathcal{T}_{k^0})$.

\item Update $\hat \beta$ and $\hat \delta$ by estimating \eqref{eq:stackedt_tilde_full} by OLS conditional on $\hat{\mathcal{T}}_{k^0}$ (and using $\bar Z(\hat{\mathcal{T}}_{k^0})$ to estimate the factors).

\item Iterate steps 3 and 4 until convergence.
\end{enumerate}	

Notice how the averages used to estimate the factors change in every step. As mentioned in Section \ref{section::assump}, if there are known factors, then these should be appended to the averages at every step.

The above algorithm is similar in spirit to the one considered by BP03 in case of a partial structural change model. While convergence to the global minimum is not guaranteed, convergence to a local optimum for the resulting iterated estimator has been shown by Sargan (1964). Tests with both simulated and real data in BP03 and here confirm that convergence is fast, typically with only one iteration needed and rarely a second.

In what follows, we prove that $\hat{\mathcal{T}}_{k^0}$ is consistent and derive its limiting distribution.

\begin{theorem}\label{thm:breakestim}
Suppose that Assumptions \ref{ass:breakdates}--\ref{ass:invert} are met. Then, for all $j=1,...,k^0+1$, as $N,\,T\to\infty$ with $T/N \to 0$,
\begin{equation}
N(\hat T_j-T_j^0) = O_p(1).
\end{equation}
\end{theorem}

As is well known, with time series data consistent estimation of the breakpoints is not possible, but only consistent estimation of the break fractions, and this is true also for BP98. By contrast, Theorem \ref{thm:breakestim} states that $\hat{\mathcal{T}}_{k^0}$ is consistent and that the rate of convergence is $N^{-1}$. The accuracy of the estimated breakpoints is therefore greatly enhanced when compared to the time series case.

Baltagi et al. (2016) consider a model that is very similar to ours but with a single break. They show that the estimated breakpoint is consistent; however, they do not provide the rate of convergence. Under stationarity, the model considered by Baltagi et al. (2017) is very similar to ours but without interactive effects and just one break. The rate given in Theorem \ref{thm:breakestim} is consistent with the one given in their Theorem 2.

Define $\xi_j = \Delta_j'\Omega_{j+1}\Delta_j/\Delta_j'\Omega_j\Delta_j$, $\phi_{1,j} = \Delta_j\Phi_j\Delta_j/\Delta_j'\Omega_j\Delta_j$ and $\phi_{2,j} = \Delta_j\Phi_{j+1}\Delta_j/\Delta_j'\Omega_{j+1}\Delta_j$, where $\Delta_j = \delta_{j+1} - \delta_{j}$, and $\Phi_j$ and $\Omega_j$ are as in Assumption \ref{ass:limit}. Let $B_{1,j}(s)$ and $B_{2,j}(s)$ be two scalar standard Brownian motions on $[0,\infty)$ that are independent of each other as well as over $j$. Also, $B_{1,j}(0)=B_{2,j}(0) = 0$. We now define $V_j(s)$ such that $V_j(s) = B_{1,j}(-s)-|s|/2$ if $s \leq 0$ and $V_j(s) = \sqrt{\xi_j\phi_{2,j}/\phi_{1,j}}B_{2,j}(s)-\xi_j |s|/2$ if $s > 0$. We now have everything we need in order to state the asymptotic distribution of $N(\hat T_j-T_j^0)$.

\begin{theorem}\label{thm:breakdist}
Suppose that Assumptions \ref{ass:breakdates}--\ref{ass:limit} are met. Then, for all $j=1,...,k^0+1$, as $N,\,T\to\infty$ with $T/N \to 0$,
\begin{equation}
\frac{(\Delta_j'\Omega_{j}\Delta_j)^2}{\Delta_j\Phi_j\Delta_j}N(\hat T_j-T_j^0) \to_w \arg \max_{s\in [0,\infty)} V_i(s).
\end{equation}
\end{theorem}

The probability density function of $\arg \max_{s\in [0,\infty)} V_i(s)$ is known analytically and is given in Bai (1997). The density function depends on $\sqrt{\xi_j\phi_{2,j}/\phi_{1,j}}$ and $\xi_j$, which can be estimated given estimates of $\Delta_j$, $\Phi_j$ and $\Omega_j$. Such an estimate can be constructed as $\hat\Delta_j = \hat \delta_{j+1} - \hat \delta_{j}$, and by specifying $\hat \Phi_j$ and $\hat \Omega_j$ analogously to $\hat \Phi$ and $\hat \Omega$, respectively, but for the estimated $j$-th subsample spanning the interval $[\hat T_{j-1}+1, \hat T_{j}]$. For example, in case of $\hat Q_j$, we take
\begin{align}
\hat \Omega_j = \frac{1}{N\Delta \hat T_j}\sum_{i=1}^N \sum_{t=\hat T_{j-1}+1}^{\hat T_{j}}\check w_{i,t}\check w_{i,t}',
\end{align}
where $\check w_{i,t}$ is as before. Once $\hat\Delta_j$, $\hat \Phi_j$ and $\hat \Omega_j$ have been obtained, we can construct $\hat \xi_j$, $\hat \phi_{1,j}$ and $\hat \phi_{2,j}$ by plugging in estimates in place of true parameters. Denote by $c_\alpha$ the $(1-\alpha/2)$-th percentile of the probability density function of $\arg \max_{s\in [0,\infty)} V_i(s)$. In analogy to Bai (1997), an asymptotically correctly sized $100(1-\alpha)$\% confidence interval for $T^0_j$ can now be constructed in the following way:
\begin{align}
\left[ \hat T_j - \left\lfloor c_\alpha \frac{\hat \Delta_j'\hat \Phi_j \hat \Delta_j }{N(\hat \Delta_j'\hat \Omega_j \hat \Delta_j)^2} \right\rfloor -1,\, \hat T_j + \left\lfloor c_\alpha \frac{\hat \Delta_j'\hat \Phi_j \hat \Delta_j }{N(\hat \Delta_j'\hat \Omega_j \hat \Delta_j)^2} \right\rfloor + 1\right].
\end{align}
Note that because the break dates are integer valued, the confidence interval is integer valued, too.

Once the presence of breaks has been established and their locations determined, $\beta$ and $\delta$ can be estimated by simply applying OLS to \eqref{eq:stackedt_tilde_full} with $\mathcal{T}_{k}$ replaced by $\hat{\mathcal{T}}_{\hat k}$.

\section{The impact of quantitative easing on bank lending}\label{section::appl}

\subsection{Motivation}

Since the 2007--2008 global financial crisis, monetary policy has been close to the zero lower bound in many countries. As a consequence, central banks have turned toward unconventional monetary policy as a means to stimulate their economies. Among several measures available, the main policy instrument has been QE. The US Federal Reserve, in particular, has implemented at least four major rounds of QE through which they purchased US Treasuries and mortgage-backed securities (MBSs) from the commercial banking sector, with the aim of boosting lending and stimulating economic activity. The most well-known are the ``QE1'', ``QE2'' and ``QE3'' rounds that took place in the aftermath of the global financial crisis. QE1 was announced in November 2008 and lasted until June 2010, and was followed by QE2, which spanned the period November 2010--June 2011. QE3 began in September 2012 and ended in October 2014. The fourth and most recent QE round is the one initiated in March 2020 as a response to the COVID--19 pandemic, henceforth labelled ``QE4''.

A common feature of QE1--QE4 is that they are massive in scale. As a reflection of this, the Federal Reserve balance sheet increased from about USD 800 billion in 2007 to over USD 8.5 trillion in 2021. This is clearly visible in Figure 1, which plots the Federal Reserve Treasury and MBS holdings over time, together with the QE dates. Because of their magnitude, the effect of these QEs on the banking sector has attracted considerable interest, so much so that there is by now a separate literature devoted to them.

\begin{center}
{\sc Insert Figure 1 about here}
\end{center}

Interestingly enough, however, while massive indeed, so far the empirical evidence regarding the effectiveness of QE has been mixed and far from conclusive. Rodnyanski and Darmouni (2017) were among the first to examine the first three QE rounds jointly. They found that banks with relatively larger holdings of MBSs expanded lending, but only during QE1 and QE3, because QE2 targeted Treasuries which were sparsely held by banks. Similar results have later been reported by Luck and Zimmermann (2020). Kapoor and Peia (2021) also study QE1--QE3. According to their results, however, only QE3 had a strong effect on liquidity creation. This last finding is in turn markedly different from that of Chakraborty et al. (2020), who document a negative effect of QE3 on commercial and industrial (C\&I) lending. For the most recent QE4 round there is to the best of our knowledge no evidence at all.

The present paper is the first to consider all four rounds. This is our first contribution. Our second contribution lies in our choice of econometric method. The standard approach in the literature is to exploit differences in exposure to QE policies across banks. The basic idea is to split the sample of banks into a treatment and a control group, where the former is assumed to be relatively more exposed to QE policy. Given that the Federal Reserve bought large quantities of mainly MBSs during the QE rounds, the argument goes on to say that banks with relatively large MBS holdings should benefit more, and hence be more exposed. The effect of QE policy is then estimated via a standard DiD regression in which banks' lending is regressed onto a dummy variable that takes on the value one whenever a bank that belongs to the treatment group has been subject to a particular QE policy and zero otherwise, control variables, and bank and time fixed effects.

While popular, the standard DiD approach to QE evaluation has (at least) two drawbacks. One drawback is that fixed effects are highly restrictive in that they require that in the absence of treatment the difference between the treatment and control groups is constant over time. This is the so-called ``parallel trend'' condition, which has attracted considerable attention in the QE literature (see, for example, Di Maggio et al., 2020, Luck and Zimmermann, 2020, and Rodnyanski and Darmouni, 2017). The reason is that banks with higher exposure to QE may be lending to firms that experience faster credit demand growth due to improvements in their borrower health. Changes in the demand for credit may therefore cause the treatment and control groups to differ systematically over time even in absence of QE. Demand-side effects is one source of non-parallel trending, but there are others, such as aggregate macroeconomic conditions and omitted variables (see, for example, Di Maggio et al., 2020, and Kapoor and Peia, 2021, for discussions). This is important because if the parallel trend condition is violated the OLS estimator is no longer consistent.

Another drawback of the DiD approach is that it requires correct specification not only of the timing of the QEs but also of the treatment and control groups, both of which are key in the construction of the treatment dummy. As mentioned earlier, existing studies focus on QE1--QE3, leaving ``Operation Twist'', the QE1 rollover, QE extensions and other sizeable purchases which the Federal Reserve conducted outside the QE rounds out of the analysis. Furthermore, the rounds were lengthy and contained periods with varying degree of asset purchases, and it is not clear if their impact began with the announcements (see Luck and Zimmermann, 2020).

If it is difficult to get the timing of the QEs right, correct specification of the treatment and control groups is literally impossible, given that strictly speaking there is no control group, since all banks are exposed to QE to some extent (see Chakraborty et al., 2020, and Luck and Zimmermann, 2020). The standard approach is to measure banks' exposure to QE by their MBS holdings, and to define the treatment and control groups as the upper and lower quantiles of the MBS distribution (prior to QE1). This raises (at least) two concerns; banks' MBS holdings are likely an imperfect measure of their QE exposure, and the choice of which quantiles to use is largely arbitrary. The treatment dummy is therefore generally mismeasured, which is a major problem, commonly referred to as ``misclassification'', because it makes the regression error correlated with the true outcome, thereby rendering OLS inconsistent (see, for example, Mahajan, 2006).

The present paper is not the first to point to these shortcomings, but it is the first to consider an econometric approach that is designed to deal with both in a rigorous way. The convention in the literature is to employ a large battery of robustness checks intended to demonstrate the validity of the conclusions. As is well known in the econometric literature, however, such checks are subject to numerous pitfalls, and can in fact be entirely misleading. One example of such a pitfall is data snooping. In the words of Lu and White (2014, page 200), ``[b]y submitting only results that may have been arrived at by specification searches designed to produce plausible results passing robustness checks, researchers can avoid having reviewers point out that this or that regression coefficient does not make sense or that the results might not be robust. And if this is enough to satisfy naive reviewers, why take a chance? Performing further analyses that could potentially reveal specification problems, such as nonlinearity or exogeneity failure, is just asking for trouble.'' But even in the absence of such data snooping, it is important to note that robustness checking is not a robust econometric approach, which in the current context means an approach that is valid in the presence of uncertainty over the validity of the parallel trend condition, the timing of the QE effects, and the specification of the treatment and control groups.

The toolbox developed in Section \ref{section::toolbox} allows for interactive effects in which there may be individual unobserved heterogeneity that changes over time as a result of common shocks. The parallel trend condition is therefore not required, which is a substantial advantage when compared to the standard DiD approach. Another advantage that we exploit in this section to identify the effect of QE on bank lending is that the toolbox can test for and date multiple structural breaks. The previous literature measures banks' exposure to QE by their MBS holdings and therefore so do we. However, instead of recoding it into a dummy variable for whether a particular bank belongs to the treatment or control group, we take the ratio as is and include it as a regressor with a potentially breaking coefficient. This makes our approach similar to those considered by Luck and Zimmermann (2020), and Kapoor and Peia (2021), in which the treatment variable is the actual MBS holdings multiplied by a dummy for each of the first three rounds of QE. The main difference is that here we treat the dates of any breaks as unknown to be estimated from the data.

\subsection{Data}

The analysis is based on bank-level quarterly data taken from the Call Reports of the Federal Deposit Insurance Corporation. The initial sample is the universe of commercial banks, observed from 2005Q3 to 2021Q3. The starting period coincides with the first Federal Reserve Treasury purchases. Following Rodnyanski and Darmouni (2017) and others, we balance our sample, keeping only those banks for which we have observations for the whole time period to control for mergers or acquisitions. Many studies in the literature aggregate their data to bank holding company level. However, whenever tested, the results for the bank level data tend to be the same (see, for example, Rodnyanski and Darmouni, 2017). In the present paper, we therefore base our analysis on this last type of data, which, in addition to being free of aggregation bias, contains a relatively large number of observations. In particular, there are $N=3,557$ banks that are observed over $T=64$ quarters, for a total of $227,648$ bank-quarter observations, which is substantially larger than many of the samples employed in the existing literature (see, for example, Chakraborty et al., 2020, and Rodnyanski and Darmouni (2017). Note also that while $T$ is large, $N$ is much larger, suggesting that our asymptotic theory based on letting $N,\,T\to\infty$ with $T/N \to 0$ should provide a very accurate approximation of the behaviour of the proposed toolbox.

We employ the same three dependent variables ($y_{i,t}$) as in Luck and Zimmermann (2020), Kapoor and Peia (2021), and Rodnyanski and Darmouni (2017). They are the logarithm of (i) total loans, (ii) real estate (RE) loans and (iii) commercial and industrial (C\&I) loans. The main regressor of interest, whose coefficient will be allowed to be breaking, is banks' MBS holdings. Unlike most existing studies, however, for reasons mentioned earlier, we do not want to rely solely on MBS holdings as a measure of banks' exposure to QE. Because of the way that they enter the model as regressors with potentially breaking coefficients, the proposed toolbox can easily accommodate additional measures. Rodnyanski and Darmouni (2017) consider Treasuries holdings as an alternative to MBS holdings. According to their results, however, this measure is relatively unimportant, which they explain by the fact that banks do not hold many Treasury securities. Chakraborty et al. (2020) therefore suggest an alternative, indirect, measure of the effect of Treasury purchases. In particular, they argue that banks with relatively high non-MBS securities holdings should benefit more from Treasury purchases lowering yields on these securities. In this section, we therefore include banks' non-MBS securities holdings as an additional measure of their exposure to QE. As usual in the literature, both measures are normalised by total assets. Also, to reduce the risk of simultaneity, following Chakraborty et al. (2020), all regressors are lagged once. Hence, in terms of the model in \eqref{eq:y}, $w_{i,t}$ is lagged MBS and non-MBS holdings over total assets.

A number of bank-level controls are included to capture differences in the scale and financial position of banks that might affect their lending activity, again following the convention in the literature. They are returns on assets (ROA), total assets, equity over total assets, cash over total assets, cost of deposits, and net income. These are the variables that go into $x_{i,t}$. GDP and inflation growth are also included to control for macroeconomic conditions and are treated as observed common factors, as discussed in Section \ref{section::assump}. The data for these last two variables are extracted from the FRED database of the Federal Reserve Bank of St. Louis.

\begin{center}
{\sc Insert Tables 1 and 2 about here}
\end{center}

Tables 1 and 2 provide a summary of each variable in the sample and some descriptive statistics, respectively. One of the descriptives included is Pesaran's (2021) CD test for the presence of cross-sectional correlation. According to the test results, the null hypothesis of no correlation is strongly rejected for all the bank-level variables, which is suggestive of common factors, and hence of non-parallel trending. Thus, as Di Maggio et al. (2020) point out, the parallel trend condition is unlikely to hold when using the type of long-span panel data considered here.

\subsection{Break testing and estimation}

We first want to test if there are any breaks present at all. Given that the QEs were of different sizes and contained different policy mixes, we do not want to assume that all of them led to breaks. This means that we want to treat the number of breaks as unknown. For this reason, we employ the $\mathrm{WDmax} F(k_{max})$ statistic, in which the null hypothesis of no breaks is tested against the alternative of up to $k_{max}$ breaks. Because the test outcome was unaffected by the weighing, we set $c_{\alpha,1}/c_{\alpha,k}=1$. We also set $k_{max} = 9$ and $\epsilon= 0.05$, which means that the critical values can be taken directly from Table I of BP98. The test values for total, RE and C\&I loans are 246.877, 97.722 and  28.671, respectively, which are all larger than the appropriate critical value at the most conservative 1\% level, 17.61.

The number of breaks turned out to be a difficult object to estimate, which is partly expected given the discussion in BP03. The main problem is, as BP98 point out, that while for relatively small values of $k$ the critical values increase markedly when an additional break is added, for $k \geq 5$ the critical values are quite flat in $k$. The effect is that if the true number of breaks is relatively large, the sequential testing procedure may detect even more breaks. As a result, the estimated number of breaks can sometimes be very large, and this is also what we find. In fact, the number of breaks is always estimated to $k_{max}$, regardless of how we set this tuning parameter. These estimates are therefore not reliable.

On the other hand, the exact choice of $k$ to use in the estimation of \eqref{eq:y} should not matter too much, as long as $k$ is chosen large enough to cover all relevant breaks. Unreported results suggest that the estimation results for $\beta$ and $\delta_1,...,\delta_{k+1}$ are quite sensitive to changes in $k$ for small values of $k$, in that the estimate of $\delta_{k+1}$ is markedly different from that of $\delta_{k}$, and the estimates of $\beta$ from one value of $k$ to another differ too. As $k$ increases, however, these differences tend to become smaller and from $k = 7$ onward they are almost completely absent. The fact that the estimated coefficients are stable after $k = 7$ suggests that with this many breaks at least the model is not underspecified. Hence, in what follows, we employ seven breaks.

\begin{center}
{\sc Insert Table 3 about here}
\end{center}

Table 3 reports the estimated breakpoints and the associated 95\% confidence interval for each of the seven breaks. The breaks are precisely estimated, having very narrow confidence intervals covering only a few quarters before and after, which is partly expected given the size of the sample. The estimated breakpoints for total and RE loans coincide, and are very similar to those for C\&I loans.

Of direct interest are the estimated break dates, which all coincide with major (QE) events. The first break takes place at about the same time as the first signs of the global financial crisis became visible. On February 27, 2007, stock prices in China and the US fell by the most since 2003 as reports of a decline in home prices and durable goods orders led to growth fears, with former chairman of the Federal Reserve Alan Greenspan predicting a recession. The second and seventh breaks coincide with the start of QE1 and QE4, respectively, which were the largest QE rounds. The third break coincides with the announcement of Operation Twist. The fourth and fifth breaks coincide with the tapering and the end of QE3, respectively. The sixth break is estimated close to the start of Federal Reserve's balance sheet normalisation programme.

Another major point about the breakpoints reported in Table 3 is that they are not the same as the conventional QE dates employed in the literature. Indeed, while the starting dates of QE1 and QE3 are within the 95\% confidence intervals for the second and fourth breaks, respectively, and the end date of QE3 is within the interval for the fifth, the end of QE1 is not included, as is the entire QE2. It therefore seems as that some of the conventional QE dates have been labelled as breaks when in fact they are not. This finding is largely consistent with studies such as Kapoor and Peia (2021), Rodnyansky and Darmouni (2017), and Luck and Zimmermann (2020), which all point to a relatively stronger impact of QE1 and QE3 on lending. More importantly, a number of the dates that we identify as breaks are not among the conventional QE dates. This means that one should interpret existing results based on the standard DiD approach with caution, since the possibility remains that they are due in part to omitted structural breaks.

\subsection{Model estimation}

With the estimated breaks in hand we turn to the estimation of \eqref{eq:y}, which we reparameterize in the following way:
\begin{align}
y_{i,t}&=x_{i,t}'\beta + \sum_{j=1}^{k_0+1}1( T_j < t \leq T_{j+1})w_{i,t}'\delta_j+e_{i,t} \notag\\
& = x_{i,t}'\beta + w_{i,t}'\delta_1 + \sum_{j=1}^{k_0}1( t> T_j)w_{i,t}'\Delta_j+e_{i,t},\label{eq:applreg2}
\end{align}
where $t=1,...,T$, $\Delta_j = \delta_{j+1} - \delta_j$ is as in Section \ref{section::breakestim} and $1(A)$ is the indicator function for the event $A$ taking the value one if $A$ is true and zero otherwise. The reason for why we use this parametrization is because at the end of a QE intervention, the Federal Reserve's balance sheet does not return to its pre-intervention levels, and therefore price effects remain. Therefore, it is not meaningful to compare say, bank lending behaviour after QE3 with bank lending behaviour before QE1 to determine the effectiveness of QE3, but rather we should compare post-QE3 behaviour to pre-QE3 behaviour. The reparameterized model in \eqref{eq:applreg2} with $\delta_1,\Delta_1,...,\Delta_{k_0}$ as opposed to $\delta_1,...,\delta_{k_0+1}$ as coefficients captures this.

According to theory, QE should cause increased lending by banks. The intuition goes as follows. The Federal Reserve purchases MBSs and Treasuries that are held by banks. Banks sell some of their holdings in these securities to the Federal Reserve, and this increases their reserves.\footnote{Increased lending brought about by increases in banks' reserves is called the ``mortgage origination channel'' by Chakraborty et al. (2020) or the ``liquidity channel'' by Darmouni and Rodnyanski (2017).} On the other hand, unsold holdings increase in value because of the increased demand.\footnote{This is what Rodnyansky and Darmouni (2017) refer to as the ``net worth channel''.} Both the increased reserves and the higher valued MBS and Treasury holdings improve the financial condition of banks, which means that they can increase their lending. Hence, in terms of \eqref{eq:applreg2}, we expect $\Delta_j$ to be positive following a QE intervention.

As outlined in Section \ref{section::modelass}, the bank-level regressors in $x_{i,t}$ and $w_{i,t}$ are augmented with their cross-section averages to account for interactive effects, and, as mentioned earlier, GDP and inflation growth are included as observed common factors. The loadings of all these factors are assumed to be time-invariant, which rules out the possibility of policy-induced breaks. In the previous literature it is quite common to allow QE to affect the overall level of lending (see, for example, Kapoor and Peia, 2021, and Rodnyansky and Darmouni, 2017). We do more. In particular, a breaking constant is included, which is equivalent to allowing for breaking bank fixed affects. This type of breaking observed factors is easily accommodated. In fact, all one has to do is to organize the breaking factors by regime similarly to $\bar W(\mathcal{T}_{k})$, and append it to $\bar Z(\mathcal{T}_{k})$. All-in-all, we include $p_x+p_w=6$ averages and three observed factors, which means that we can allow for up to nine common factors where one is potentially breaking. This should be more than enough to capture the unobserved heterogeneity of bank lending.

\begin{center}
{\sc Insert Table 4 about here}
\end{center}

The estimation results for each loan category are reported in Table 4.\footnote{The Stata command \texttt{xtdcce2} by Ditzen (2018) was used to obtain the regression results.} We begin by considering the results for total loans. The effect of banks' MBS and non-MBS holdings tend to have the same sign, which we interpret as that Federal Reserve's purchases of MBSs and Treasuries work in the same direction. We also see that bank lending is generally not affected by the exact holding mix. One exception is in the first and second regimes, which coincide with the pre-crisis and financial crisis periods, respectively. In particular, while the estimated effects are of the same sign, only banks' non-MBS holdings enter significantly. The fact that it is only banks' non-MBS holdings that matters is expected given Figure 1. The estimated effect of these holdings is positive in the first regime (that is, $\delta_1$ is estimated to be positive), suggesting that increased Treasury purchases by the Federal Reserve caused banks to increase lending, which is just as expected given the discussion of the previous paragraph. But then in the second regime the sign changes to negative (that is, $\Delta_1$ is estimated to be negative), which means that the estimated effect of banks' non-MBS holdings on their lending decreases significantly when compared to the first regime. This last effect can be explained by considering the health of the banking sector, as we describe in detail below. The effect of banks' MBS holdings on lending decreases too, although not significantly so. The holding mix therefore matters here. From this point on, however, the estimated effect is basically the same regardless of the type of holding being considered, suggesting that the mix is unimportant. We also see that for quite some time following the crisis the estimated effect of banks' MBS and non-MBS holdings on lending is either insignificant or significantly negative, suggesting that the impact of Federal Reserve's purchases of MBSs and Treasuries on lending is either the same as during the crisis or even more negative. This development is brought to a halt at the time of the fifth break (the end of QE3) when the estimated effect of banks' MBS and non-MBS holdings on lending increases significantly when compared to previous quarters, which is consistent with the results of Luck and Zimmermann (2020). This increase continues until the end of the sample period.

As a measure of the overall effect of QE on lending, we look at the difference between the first (pre-intervention) regime and the last (QE4) regime, as given by $\sum_{j=1}^{k_0}\Delta_j = \delta_{k_0+1} - \delta_1$. The overall effect for MBS and non-MBS holdings are estimated to $0.165$ and $0.007$, respectively. This is when we ignore coefficient estimates that are insignificant. If we include all estimates, said effects are $-0.118$ and $-0.115$, respectively. Hence, even if in the latter part QE seems to have worked, considering the full sample period QE has not been very effective in spurring credit flow.

\begin{center}
{\sc Insert Figure 2 about here}
\end{center}

In order to ease the interpretation of the results reported in Table 4, in Figure 2 we plot the average Tier 1 risk based capital ratio, which is an indicator of banks' financial strength and is regulated by Basel III. Between 2008Q2 and 2014Q4 this ratio is increasing steeply, suggesting that banks retain some of the reserves produced by Federal Reserve's purchases in order to improve their financial position. This explains why in this period lending does not increase. The Tier 1 ratio flattens after 2014Q4, only to increase again in 2020Q1. The fact that in the 2014Q4--2020Q1-period the Tier 1 ratio is relatively stable means that banks are no longer building up reserves to the same extent as before. Banks therefore have more room to expand lending, and this is reflected in the estimation results. The observed increase in the Tier 1 ratio after 2020Q1 is compensated in part by the scale of Federal Reserve's purchases during QE4. Banks are therefore able to expand lending while at the same time cover their capitalisation needs. The importance of banks' financial position for their ability to lend is consistent with studies such as Kim and Sohn (2017).

The above conclusions for total loans apply also to RE loans. In fact, the estimation results are almost identical, except that the contractionary effect of the crisis is more pronounced for RE than for total loans. The results for C\&I loans are quite different, though. Note in particular how in the second (financial crisis) regime banks' MBS and non-MBS holdings are no longer significant, and that the positive effect in the sixth (end of QE3) regime is much stronger than for total and RE loans. This last difference corroborates the finding of Luck and Zimmermann (2020) that C\&I lending has not been very responsive to QE, except post-QE3 when it increases. It also strengthens our conclusion that it is mainly from the fifth break onwards that Federal Reserve's interventions have lead to increased lending. The overall effects of MBS and non-MBS holdings on C\&I lending are estimated to $0.032$ and $0.701$, respectively, suggesting that while Federal Reserve's MBS purchases have again not been very effective, its Treasury purchases has had a substantial effect, which consistent with the findings of Chakraborty et al. (2020). It is also consistent with the Figure 1 and the massive increase in Federal Reserve's Treasuries holdings during QE4.

\section{Conclusions}\label{section::concl}

This paper provides a toolbox that meets the needs of researchers interested in a linear panel data model with interactive effects and possibly multiple structural breaks. The toolbox allows researchers to test for the presence of breaks, and, if breaks are detected, to estimate the location of the breaks and construct confidence intervals for the true breakpoints.

The new toolbox is employed to study the effects of the US Federal Reserve's QE interventions during 2005--2021 on bank lending. The idea behind QE is that banks should transform the reserves generated by the interventions into loans, which should in turn increase consumption and employment. However, since the composition of banks' balance sheets affect the amount of loans they create, the impact of QE on lending is not a priori clear. Our results suggest that while QE has led to a number of breaks in bank's lending behaviour, it is only towards the end of the sample period that lending has increased significantly.

\section*{References}

\begin{description}	
\item Andrews, D. W. K. (1993). Tests for Parameter Instability and Structural Change with Unknown Change Point. \emph{Econometrica} \textbf{61}, 821--856.

\item Antoch, J., J. Hanousek, L. Horv\'{a}th, M. Hu\v{s}kov\'{a} and S. Wang (2019). Structural Breaks in Panel Data: Large Number of Panels and Short Length Time Series. \emph{Econometric Reviews} \textbf{38}, 828--855.

\item Bai, J. (1997). Estimation of a Change Point in Multiple Regression Models. \emph{Review of Economics and Statistics} \textbf{79}, 551--563.

\item Bai, J. (1999). Likelihood Ratio Tests for Multiple Structural Changes. \emph{Journal of Econometrics} \textbf{91}, 299--323.

\item Bai, J. (2009). Panel Data Models with Interactive Fixed Effects. \emph{Econometrica} \textbf{77}, 1229--1279.

\item Bai, J. (2010). Common Breaks in Means and Variances for Panel Data. \emph{Journal of Econometrics} \textbf{157}, 78--92.

\item Bai, J., and P. Perron (1998). Estimating and Testing Linear Models with Multiple Structural Changes. \emph{Econometrica} \textbf{66}, 47--78.

\item Bai, J., and P. Perron (2003a). Critical Values for Multiple Structural Change Tests. \emph{Econometrics Journal} \textbf{6}, 72--78.

\item Bai, J., and P. Perron (2003b). Computation and Analysis of Multiple Structural Change Models. \emph{Journal of Applied Econometrics} \textbf{18}, 1--22.

\item Baltagi, B. H., Q. Feng and C. Kao (2016). Estimation of Heterogeneous Panels with Structural Breaks. \emph{Journal of Econometrics} \textbf{191}, 176--195.

\item Baltagi, B. H., C. Kao and L. Liu (2017). Estimation and Identification of Change Points in Panel Models with Nonstationary or Stationary Regressors and Error Term. \emph{Econometric Reviews} \textbf{36}, 85--102.

\item Baltagi, B. H., C. Kao and F. Wang (2021). Estimating and testing high dimensional factor models with multiple structural changes. \emph{Journal of Econometrics} \textbf{220}, 349--365.

\item Breitung, J., and S. Eickmeier (2011). Testing for Structural Breaks in Dynamic Factor Models. \emph{Journal of Econometrics} \textbf{163}, 71--84.

\item Bryan, K. A., and Y., Ozcan (2021). The Impact of Open Access Mandates on Invention. \emph{The Review of Economics and Statistics} \textbf{103} (5) 954-–967.

\item Boldea, O., B. Drepper and Z. Gan (2020). Change Point Estimation in Panel Data with Time-Varying Individual Effects. \emph{Journal of Applied Econometrics} \textbf{35}, 712--727.

\item Chakraborty, I., I. Goldstein and A. MacKinlay (2020). Monetary Stimulus and Bank Lending. \emph{Journal of Financial Economics} \textbf{136}, 189–218.

\item Chen, B., and L. Huang, (2018). Nonparametric testing for smooth structural changes in panel data models. \emph{Journal of Econometrics} \textbf{202}, 245--267.

\item Di Maggio, M., A. Kermani and C. J. Palmer (2020). How Quantitative Easing Works: Evidence on the Refinancing Channel. \emph{Review of Economic Studies} \textbf{87}, 1498--1528.

\item Ditzen, J. (2018). xtdcce2: Estimating Dynamic Common Correlated Effects in Stata. \emph{Stata Journal} \textbf{18}, 585--617.

\item Ditzen, J., Y. Karavias and J. Westerlund (2022). Testing and Estimating Structural Breaks in Time Series and Panel Data in Stata. arXiv:2110.14550 [econ.EM]

\item Feng, G., J. Gao, B. Peng and X. Zhang (2017). A varying-coefficient panel data model with fixed effects: Theory and an application to US commercial banks. \emph{Journal of Econometrics} \textbf{196}, 68--82.

\item Kaddoura, Y., and J. Westerlund (2022). Estimation of Panel Data Models with Random Interactive Effects and Multiple Structural Breaks when $T$ is Fixed. Forthcoming in \emph{Journal of Business \& Economic Statistics}.

\item Kapoor, S., and O. Peia (2021). The Impact of Quantitative Easing on Liquidity Creation. \emph{Journal of Banking and Finance} \textbf{122}, 105998.

\item Karavias, Y., P. Narayan and J. Westerlund (2022). Structural Breaks in Interactive Effects Panels and the Stock Market Reaction to COVID--19. Forthcoming in \emph{Journal of Business \& Economic Statistics}.

\item Kim, D., and W. Sohn (2017). The Effect of Bank Capital on Lending: Does Liquidity Matter? \emph{Journal of Banking \& Finance} \textbf{77}, 95--107.

\item Li, D., J. Qian and L. Su (2016). Panel Data Models With Interactive Fixed Effects and Multiple Structural Breaks. \emph{Journal of the American Statistical Association} \textbf{111}, 1804--1819.

\item Lu, X., and H. White (2014). Robustness Checks and Robustness Tests in Applied Economics. \emph{Journal of Econometrics} \textbf{178}, 194--206.

\item Luck, S., and T. Zimmermann (2020). Employment Effects of Unconventional Monetary Policy: Evidence from QE. \emph{Journal of Financial Economics} \textbf{135}, 678–703.

\item Mahajan, A. (2006). Identification and Estimation of Regression Models with Misclassification. \emph{Econometrica} \textbf{74}, 631–665.

\item Pesaran, M. H. (2006). Estimation and Inference in Large Heterogeneous Panels with a Multifactor Error Structure. \emph{Econometrica} \textbf{74}, 967--1012.

\item Pesaran, H. M. (2021). General Diagnostic Tests for Cross-Sectional Dependence in Panels. Forthcoming in \emph{Empirical Economics}.

\item Qian, J., and L. Su (2016). Shrinkage Estimation of Common Breaks in Panel Data Models via Adaptive Group Fused Lasso. \emph{Journal of Econometrics} \textbf{191}, 86--109.

\item Rodnyansky, A., and O. Darmouni (2015). The Effects of Quantitative Easing on Bank Lending Behavior. \emph{Review of Financial Studies} \textbf{37}, 3858--3887.

\item Sargan, J. D. (1964). Wages and Prices in the United Kingdom: A Study in Econometric Methodology. In Hart, P. E., G. Mills and J. K. Whitaker (Eds), \emph{Econometric Analysis for National Economic Planning}, 25--54. Butterworths: London.

\item Stock, J. H., and M. W. Watson (1996). Evidence on Structural Stability in Macroeconomic Time Series Relations. \emph{Journal of Business \& Economic Statistics} \textbf{14}, 11--30.

\item Stock, J. H., and M. W. Watson (2003). Forecasting Output and Inflation: The Role of Asset Prices. \emph{Journal of Economic Literature} \textbf{XLI}, 788--829.

\item Westerlund, J., and J.-P. Urbain (2015). Cross-Sectional Averages versus Principal Components. \emph{Journal of Econometrics} \textbf{185}, 372--377.

\item Westerlund, J. (2018). CCE in Panels with General Unknown Factors. \emph{Econometrics Journal} \textbf{21}, 264--276.

\item Westerlund, J. (2019). On Estimation and Inference in Heterogeneous Panel Regressions with Interactive Effects. \emph{Journal of Time Series Analysis} \textbf{40}, 852--857.

\item Zhu, H., V. Sarafidis and M. J. Silvapulle (2020). A New Structural Break Test for Panels with Common Factors. \emph{Econometrics Journal} \textbf{23}, 137--155.
\end{description}

\clearpage

\begin{figure}%
	\caption{Federal Reserve's MBS and Treasury holdings and QE rounds.}%
	\includegraphics[width=16cm,keepaspectratio]{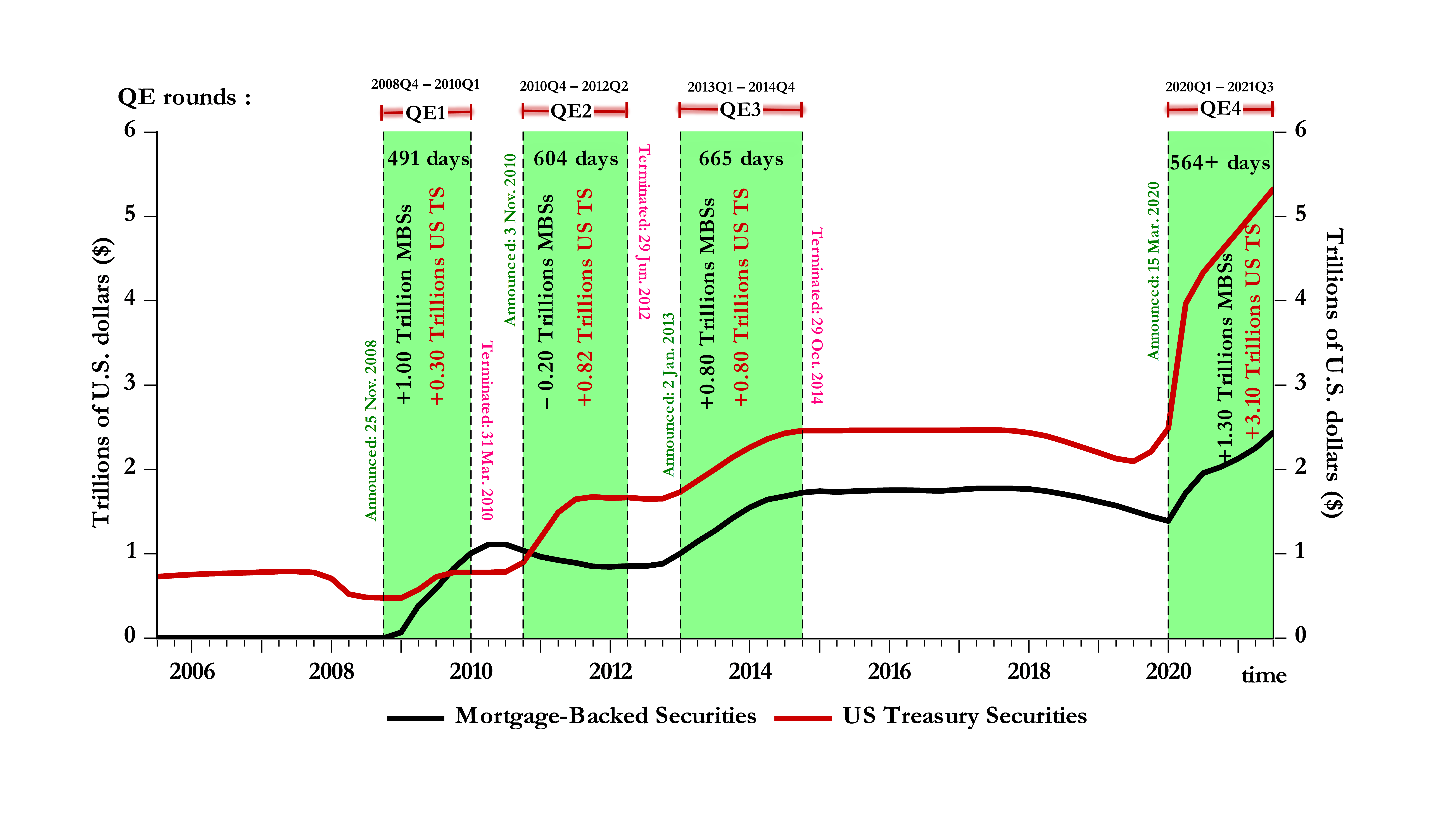}
	\label{Fig:1}
\end{figure}

\begin{table}[]
	\caption{Variable definitions.}
	\hskip 12 pt
	\centering
	\footnotesize
	\begin{tabular}{ll}\hline\hline
		Variable         & Definition                                                    \\
		\hline
		Dependent variables         &                                                    \\
		~~ Total loans      & Log of total loans                                      \\
		~~ RE loans         & Log of real estate loans                                \\
		~~ C\&I loans       & Log of commercial and industrial loans                    \\
		Regressors         &                                                    \\
		~~ MBS     & Mortgage-backed securities over total assets                  \\
		~~ non-MBS       & Total securities held minus MBS securities over total assets \\
		~~ Total Assets     & Log of total assets                                     \\
		~~ Equity           & Equity over total assets                                      \\
		~~ ROA              & Return on assets                                              \\
		~~ Bank cash        & Balance sheet cash flow over total assets                     \\
		~~ Cost of deposits & Interest expense of deposits over total assets                \\
		~~ Net income       & Net income over total assets      \\
		~~ Inflation       	 & US inflation growth      \\
		~~ GDP       	 & US per capita GDP growth        \\ \hline\hline
	\end{tabular}
	\begin{tablenotes}
		\item \emph{Notes}: All regressors are lagged one period. The data are sourced from Federal Deposit Insurance Corporation and FRED.																									\end{tablenotes}
	\label{Tab:def}
\end{table}

\begin{table}[]
	\caption{Descriptive statistics.}
	\hskip 12 pt
	\centering
	\footnotesize
	\begin{tabular}{lccccc} \hline\hline
		Variable         & Mean   & STD & Min    & Max    & CD         \\ \hline
		Total loans      & 11.709 & 1.482     & 6.590  & 20.773 & 10124.055*** \\
		RE loans          & 11.273 & 1.610     & 3.638  & 20.008 & 9489.100***    \\
		C\&I loans        & 9.508  & 1.709     & 0.693  & 19.684 & 5728.323***  \\
		MBS     & 0.075  & 0.092     & 0.000  & 0.820  & 623.285***   \\
		non-MBS        & 0.161  & 0.125     & 0.000  & 0.926  & 2089.694***  \\
		Total assets     & 12.228 & 1.399     & 8.004  & 21.914 & 14258.548*** \\
		Equity           & 0.111  & 0.032     & -0.026 & 0.730  & 2095.189***  \\
		ROA              & 0.103  & 0.082     & -3.083 & 2.548  & 1593.754***  \\
		Bank cash        & 0.082  & 0.081     & 0.000  & 0.939  & 4513.740***   \\
		Cost of deposits & 0.006  & 0.006     & 0.000  & 0.123  & 19083.858*** \\
		Net income       & 0.006  & 0.006     & -0.209 & 0.113  & 9888.248***  \\
		Inflation & 0.002  & 0.770     & -3.847 & 1.931  &  --            \\
		GDP       & 0.887  & 1.792     & -9.793 & 8.184  &  -- \\ \hline\hline
	\end{tabular}
	\begin{tablenotes}																					
		\item  \emph{Notes}: ``Mean'', ``STD'', ``Min'' and ``Max'' refer to the sample average, the sample standard deviation, the sample minimum value and the sample maximum value. The column labelled ``CD'' reports the results obtained by applying Pesaran's (2021) test for cross-sectional correlation. The CD test results for GDP and inflation growth are not reported as these variables do not vary by bank. The superscripts ``*'', ``**'' and ``***'' denote statistical significance at the $10$\%, $5$\% and $1$\% levels, respectively.
	\end{tablenotes}
\label{Tab:descr}
\end{table}

\begin{table}[]
	\caption{Estimated break dates and 95\% confidence intervals.}
	\hskip 12 pt
	\centering
	\footnotesize
	\begin{tabular}{ccccccccc}\hline\hline
		& \multicolumn{2}{c}{Total loans}            &            & \multicolumn{2}{c}{RE loans}          &           & \multicolumn{2}{c}{C\&I loans}                     \\\cline{2-3}\cline{5-6}\cline{8-9}
		Break & Date   & 95\% CI      &       & Date   & 95\% CI         &    & Date   & 95\% CI             \\ \hline
		1     & 2007Q1 & {[}2006Q4, 2007Q2{]}& & 2007Q1 & {[}2006Q4, 2007Q2{]}& & 2007Q1 & {[}2006Q4, 2007Q2{]} \\
		2     & 2009Q1 & {[}2008Q4, 2009Q2{]}& & 2009Q1 & {[}2008Q4, 2009Q2{]}& & 2009Q1 & {[}2008Q4, 2009Q2{]} \\
		3     & 2011Q3 & {[}2011Q2, 2011Q4{]}& & 2011Q3 & {[}2011Q2, 2011Q4{]}& & 2011Q3 & {[}2011Q2, 2011Q4{]} \\
		4     & 2013Q3 & {[}2013Q2, 2013Q4{]}& & 2013Q3 & {[}2013Q2, 2013Q4{]}& & 2013Q3 & {[}2013Q2, 2013Q4{]} \\
		5     & 2014Q4 & {[}2014Q3, 2015Q1{]}& & 2014Q4 & {[}2014Q3, 2015Q1{]}& & 2014Q4 & {[}2014Q3, 2015Q1{]} \\
		6     & 2017Q1 & {[}2016Q4, 2017Q2{]}& & 2017Q1 & {[}2016Q4, 2017Q2{]}& & 2016Q4 & {[}2016Q3, 2017Q1{]} \\
		7     & 2020Q1 & {[}2019Q4, 2020Q2{]}& & 2020Q1 & {[}2019Q4, 2020Q2{]}& & 2020Q1 & {[}2019Q4, 2020Q2{]}\\ \hline\hline
	\end{tabular}
	\begin{tablenotes}																					
		\item  \emph{Notes}: ``Date'' and ``95\% CI'' refer to the estimated breakpoint and the 95\% confidence interval, respectively, for each of the seven breaks.
	\end{tablenotes}
\label{Tab:breakdates}
\end{table}

\begin{table}[]
	\caption{Regression results.}
	\hskip 12 pt
	\centering
	\footnotesize
	\begin{tabular}{cccccccccc}
		\hline\hline
& &\multicolumn{2}{c}{Total loans}   &     &\multicolumn{2}{c}{RE loans}   &     &\multicolumn{2}{c}{C\&I loans}        \\\cline{3-4}\cline{6-7}\cline{9-10}
COEF		& & EST & SE & & EST & SE & & EST & SE\\
		\hline
\multicolumn{10}{c}{Regressor with breaking coefficient: MBS} \\
$\delta_1$ &  &     0.180         &    0.209& &      0.399\sym{**} &     0.161& &      0.113         &     0.411\\
$\Delta_1$ &  &    -0.293         &    0.215&  &    -0.572\sym{***}&     0.168& &     -0.408         &     0.425\\
$\Delta_2$ &  &   -0.0233         &    0.053& &    -0.046         &    0.059&   &    0.291\sym{**} &     0.125\\
$\Delta_3$ &  &    -0.192\sym{***}&    0.043&  &    -0.185\sym{***}&    0.048& &     -0.197\sym{*}  &     0.104\\
$\Delta_4$ &  &    -0.209\sym{***}&    0.057&   &   -0.227\sym{**} &    0.089&  &    -0.555\sym{***}&     0.139\\
$\Delta_5$ &  &     0.329\sym{***}&    0.059&  &     0.354\sym{***}&    0.100&  &      0.782\sym{***}&     0.165\\
$\Delta_6$ &  &    0.033         &    0.053&   &    0.121         &    0.074& &     -0.289\sym{**} &     0.142\\
$\Delta_7$ &  &     0.237\sym{***}&    0.085&   &    0.166         &     0.103&  &     0.209         &     0.199\\
\multicolumn{10}{c}{Regressor with breaking coefficient: non-MBS}   \\
$\delta_1$ &  &     0.122\sym{**} &    0.051&  &    0.092         &    0.060&  &   -0.100         &     0.114\\
$\Delta_1$ &  &  -0.223\sym{***}&    0.060&  &    -0.299\sym{***}&    0.073& &   -0.003         &     0.134\\
$\Delta_2$ &  &  -0.060         &    0.040&  &   0.010         &    0.049&  &    -0.137         &    0.091\\
$\Delta_3$ &  & -0.074\sym{**} &    0.032&  &     -0.101\sym{***}&    0.038&   &  -0.033         &    0.082\\
$\Delta_4$ &  &  -0.259\sym{***}&    0.042&  &    -0.219\sym{***}&    0.056&   &   -0.213\sym{*}  &     0.112\\
$\Delta_5$ &  &    0.273\sym{***}&    0.059&  &     0.273\sym{***}&    0.076&   &    0.424\sym{***}&     0.139\\
$\Delta_6$ &  & -0.063         &    0.060&  &   -0.059         &    0.071&    &  -0.160         &     0.129\\
$\Delta_7$ &  &   0.290\sym{***}&    0.055&  &     0.204\sym{***}&    0.064&    &   0.490\sym{***}&     0.132\\
		\hline
\multicolumn{2}{l}{Interactive effects} &   \multicolumn{2}{c}{Yes} &  &   \multicolumn{2}{c}{Yes}&   &   \multicolumn{2}{c}{Yes}\\
\multicolumn{2}{l}{Observed factors} &      \multicolumn{2}{c}{Yes} &  &     \multicolumn{2}{c}{Yes} &   &    \multicolumn{2}{c}{Yes}\\
\multicolumn{2}{l}{Bank-level controls} &   \multicolumn{2}{c}{Yes} &  &   \multicolumn{2}{c}{Yes} &    &  \multicolumn{2}{c}{Yes} \\
		\hline\hline
	\end{tabular}
	
	\begin{tablenotes}																					
		\item  \emph{Notes}: The table reports the results obtained by fitting \eqref{eq:applreg2} by CCE while allowing the coefficients of both MBS and non-MBS to be breaking. ``COEF'', ``EST'' and ``SE'' refer to the estimated coefficient, the associated point estimate and its standard error, respectively. Here $\delta_1$ is the coefficient of the first regime, while $\Delta_j = \delta_{j+1} - \delta_j$ is the change in the coefficient from regime $j$ to $j+1$. All models are fitted with cross-sectional averages of the regressors to account for unknown interactive effects, and fixed effects, GDP and Inflation as observed common factors. The fixed effects are allowed to be breaking, while the coefficients of GDP and Inflation are not. The included bank-level controls are Total assets, Equity, ROA, Bank cash, Cost of deposits, and Net income. The superscripts ``*'', ``**'' and ``***'' denote statistical significance at the $10$\%, $5$\% and $1$\% levels, respectively.
	\end{tablenotes}
	\label{Tab:regres}
\end{table}

\begin{landscape}
\begin{figure}%
	\center
	\caption{Bank lending, the Tier 1 risk based capital ratio and the estimated structural breaks.}%
	\includegraphics[width=22cm,keepaspectratio]{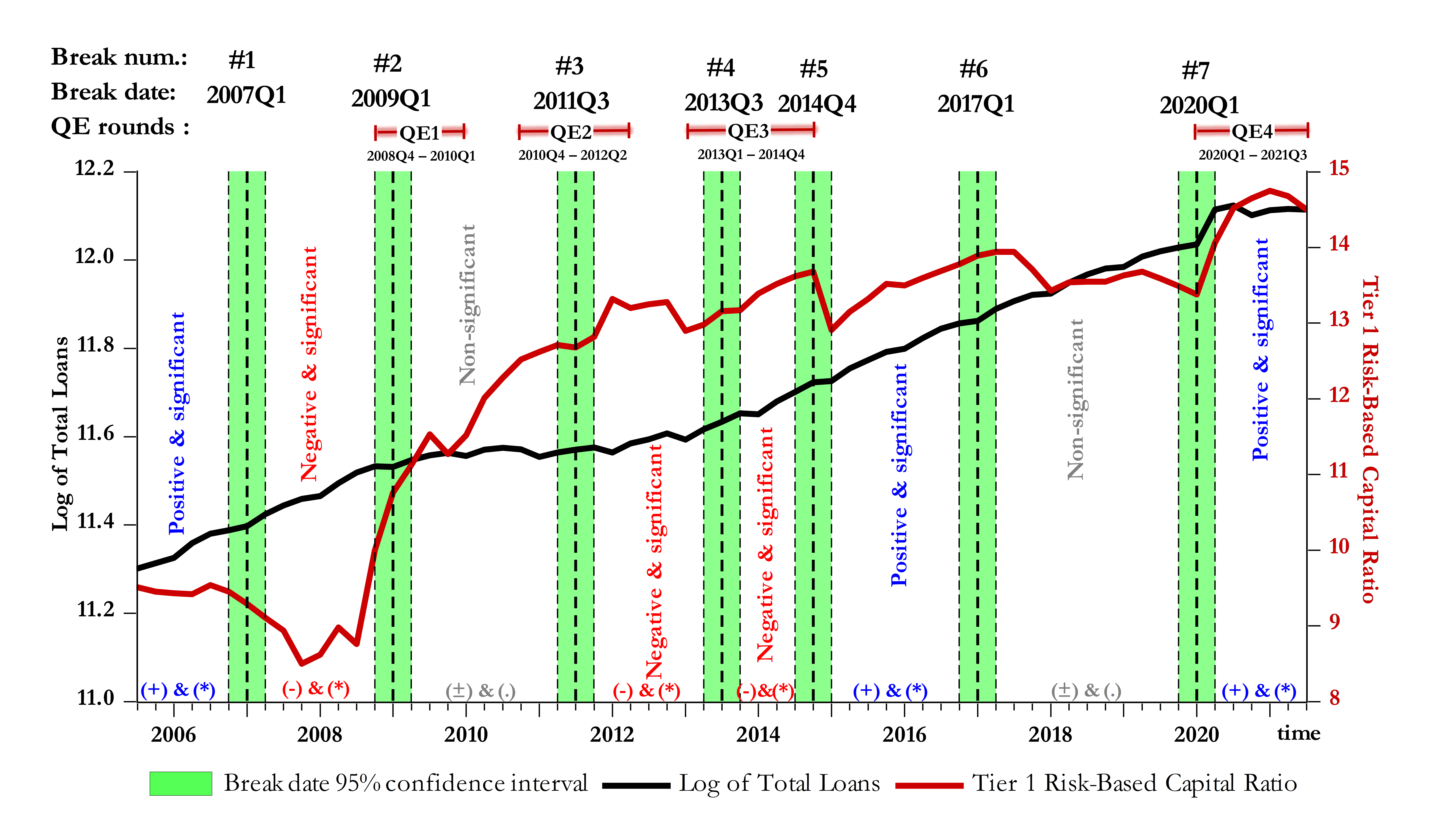}%
	\caption*{\footnotesize \emph{Notes:} The figure plots the average log total loans, the estimated breakpoints and their 95\% confidence intervals, and the Tier 1 risk based capital ratio. The red, blue and grey colored text in the background of each estimated regime indicates the sign and statistical significance of the estimated coefficients. Grey colored text: MBS and non-MBS securities holdings have statistically insignificant coefficient estimates. Red (blue) colored text: At least one of the estimated coefficients of MBS or non-MBS securities holdings have a negative (positive) and significant coefficient.}
	\label{Tab:2}
\end{figure}
\end{landscape}

\end{document}